\documentclass[a4paper,12pt]{article}
\usepackage[pdftex]{graphicx}
\usepackage{graphicx}
\usepackage{setspace}
\usepackage{indentfirst}
\usepackage{subfigure}
\usepackage{amssymb}
\usepackage{amsmath}
\usepackage{xcolor}
\usepackage[noadjust]{cite}
\usepackage{authblk}
\usepackage{afterpage}
\usepackage[square,comma,sort&compress]{}
\usepackage{sectsty}
\sectionfont{\fontsize{12}{12}\selectfont}  
\subsectionfont{\fontsize{12}{12}\selectfont}  
\font\myfont=cmr12 at 12pt
\author[]{\myfont Ashish Kumar Singh}
\author[]{\myfont Awaneesh Singh\thanks{awaneesh11@gmail.com}}
\affil[]{\myfont Department of Physics, Indian Institute of Technology (BHU), Varanasi, Uttar Pradesh-221005, India.}
\title{\fontsize{12}{12} \bf{Phase separation kinetics of block copolymer melts confined under moving parallel walls: a DPD study}}
\date{}

\begin{document}
\onehalfspacing
\maketitle

\begin{abstract}
We use dissipative particle dynamics (DPD) simulations to study the effect of shear on domain morphology and kinetics of microphase separating critical diblock copolymer (BCP) bulk melts. The melt is confined within two parallel solid walls at the top and bottom of the simulation box. The shear is induced by allowing the walls to move in a direction with a specific velocity. We explore the following cases: (i) walls are fixed, (ii) only the top wall moves, (iii) both walls move in the same direction, and (iv) both walls move in opposite directions. After the temperature quench, we monitor the effect of shear on evolution morphology, the scaling behavior of the system, and the characteristic length scale and growth. The characteristic length scale follows typical power-law behavior at early times and saturates at late times when both walls are fixed. The length scale changes significantly with shear caused by wall velocities. The usual lamellar morphology, which is not achieved for case 1 within the considered simulation time steps, is noticed much earlier for the nonzero wall velocity cases. Specifically, it is seen much before in case 4 than in the other cases. We find that the shear viscosity decreases (shear-thinning) with wall velocity (shear rate) for all the cases at a given coarsening time. Overall, we report the influence rule of shear rates on microphase separation kinetics of BCP melts. This study can provide a scheme to anticipate and design anisotropic microstructures under the application of externally controlled wall shear that may further guide in producing the various composite materials with superior mechanical and physical properties.
\end{abstract}

\newpage
\section{Introduction}
\label{Intro}
Microphase separation of block copolymer (BCP) melts \cite{leibler1980theory, hashimoto1980domain, ohta1986equilibrium, bates1990block, hamley2009ordering} has continued to be an active area of research for decades \cite{singh2015kinetics, gavrilov2013phase, nebouy2020coarse, stuparu2012phase, beardsley2019computationally}. BCP is a long linear chain molecule ($A_nB_m$) composed of two covalently bonded incompatible sub-chains of $n$ $A$- and $m$ $B$-monomer beads. When a homogeneous BCP melt in its high-temperature homogeneously mixed stable phase is quenched below the critical temperature, the system becomes thermodynamically unstable. The small fluctuations in the density field grow, and the phase separation occurs, forming $A$-rich and $B$-rich phases \cite{binder1994phase, bray2002theory, puri2009kinetics}. Since the chemically bonded immiscible sequences cannot be separated indefinitely, their sub-chain lengths control the spatial scale. Thus, coarsening gives rise to morphologies at a microscopic length scale, i.e., micro-domains rich in either of the two components are formed. Different morphologies like lamellae, gyroids, cylinders, and spheres depend on the relative compositions ($n:m$) of sub-chains \cite{leibler1980theory, hashimoto1980domain, ohta1986equilibrium, bates1990block}.   

Domain coarsening is a well-known scaling phenomenon where the characteristic length scale, $R(t)$ follows a power-law equation, $R(t)\sim t^\phi $ asymptotically \cite{binder1994phase, bray2002theory, puri2009kinetics}. Here, $\phi$ is the growth exponent, which depends on the relevant coarsening mechanism \cite{bray2002theory}. BCP melts coarsen into the well-ordered structures differently than macrophase separation of polymer blends.\cite{singh2014kinetics, singh2015phase, singh2021photo, singh2018role}. Numerous theoretical \cite{leibler1980theory, hashimoto1980domain, ohta1986equilibrium, bates1990block} and simulation \cite{singh2015kinetics, liu1989dynamics} efforts have demonstrated that the variation of characteristic length scale with time initially follows a power-law with $\phi \sim 1/4$ in the weak-segregation regime\cite{leibler1980theory} and $\phi \sim 1/3$ in the strong-segregation regime \cite{hashimoto1980domain, ohta1986equilibrium, liu1989dynamics, singh2015kinetics}. The segregating BCP melts evolve into frozen microstructures with various morphologies in the asymptotic limit. It is further demonstrated that the effects of hydrodynamics only accelerate the growth kinetics at pre-asymptotic stages compared to diffusive dynamics \cite{maurits1998hydrodynamic, groot1999role, groot1998dynamic, shiwa2000hydrodynamic, xu2005scaling}. Similar frozen morphologies are observed at late times, as in the diffusive case \cite{maurits1998hydrodynamic, groot1999role, groot1998dynamic, shiwa2000hydrodynamic, xu2005scaling}.

The self-assembled BCP morphologies are the topic of vital interest due to their rapidly increasing applications in various fields such as advanced materials \cite{feng2017block}, lithography \cite{feng2017block, stoykovich2006block, harrison2004lithography}, drug delivery \cite{feng2017block, rosler2012advanced, o2006cross}, and porous materials \cite{feng2017block, ma2013direct}. The latest applications of BCPs come from their self-assembly on the nanoscale \cite{fraser2006orientation}; these properties rely on the long-range order of BCP melts \cite{ahmed2003magnetic, thurn2000ultrahigh}. The change in morphology considerably affects the final properties of multiphase materials. Therefore, coarsening domain morphology can significantly influence the final material properties. 

The control over evolved morphology at the molecular scale can help develop new and robust materials with great industrial applications. Various methods such as shear \cite{hamley2001structure}, electric field \cite{morkved1996local}, chemically patterned substrate \cite{ouk2003epitaxial}, etc., have been proposed in the recent past to speed up BCP melts' segregation. Using either steady or oscillatory shear \cite{arya2005shear} is the more straightforward approach. Their application to BCPs plays a dominant role in determining their self-assembled structure. Using these methods, spherical, cylindrical, and lamellar morphologies have been illustrated successfully \cite{almdal1993dynamically, mcconnell1995long, scott1992shear, koppi1993shear, winey1993interdependence, chen1997dynamics}. Some recent works on BCP thin films \cite{angelescu2004macroscopic, angelescu2005shear} have improved the alignment of cylindrical and spherical domain evolution using shear; it is believed that microphases try to minimize the effect of shear by reorienting themselves towards shear flow \cite{arya2005shear}. 

Nevertheless, most of these studies were focused mainly on either $2d$ surfaces\cite{hamley2001structure, arya2005shear} such as thin substrates/films\cite{angelescu2004macroscopic, angelescu2005shear} or studying the equilibrium properties\cite{nikoubashman2013, peters2012} in BCP melts for a small system sizes. \cite{almdal1993dynamically, mcconnell1995long, scott1992shear, koppi1993shear, winey1993interdependence, chen1997dynamics} Thus, a detailed DPD simulation study of the kinetics of microphase separation of BCP bulk melts under the influence of shear is still lacking. In contrast to any coarse-grained models, which use uncontrolled approximations to model the velocity field, the DPD approach has the considerable advantage of naturally incorporating flow fields and the hydrodynamic effects in the system.\cite{singh2015kinetics} Furthermore, the influence of physical boundary and the competition between shear and phase separation kinetics in BCP melts are crucial in understanding the morphology and related properties changes \cite{arya2005shear}. Thus, a more detailed study on growth kinetics, particularly the growth exponent, scaling properties, and structural anisotropy due to variation in shear rate, are of paramount importance for a BCP melt under confinement. The shear is introduced by moving parallel walls confining the system from the top and the bottom. Herein, we choose a symmetric BCP melt where each chain has two incompatible sub-chains of equal length. The solid walls have the same interaction with each sub-chain; thus, both phases experience an equal amount of shear caused by the wall velocity in a particular direction. Therefore, it is natural to inquire about the shear effect on the microphase separation kinetics in BCP melts. We ask the following questions: does shear stimulate or delay the segregation process in the bulk system? Further, do micro-scale morphologies vary with low and high wall velocities? Finally, what would be the effect on scaling functions and growth laws?

In the next section, we briefly describe the DPD methodology that captures the effect of shear on the phase separation kinetics of BCP melts. Then, the simulation results are discussed in Section \ref{Nr}. Finally, we conclude the paper in Section \ref{sumcon}.

\section{Methodology}
\label{Method}
\subsection{Dissipative particle dynamics}
\label{DPD}
DPD is a powerful simulation technique to study the kinetic behavior of complex soft mesoscale systems \cite{singh2021photo,groot1997dissipative,espanol1995statistical,espanol2017perspective}. In DPD, a cluster of particles or molecules is modeled as a single bead, which makes DPD a more thriving numerical tool to simulate the system over a higher length and time scale than a traditional molecular dynamics (MD) simulation technique \cite{groot2006local,nikunen2003would,nikunen2007reptational}. We integrate Newton's equation of motion for the system's time evolution:
\begin{equation}
\label{Newton}
\frac{d\vec{p}_i}{dt}=\vec{f}_i(t)= \sum_{i \neq j} \vec{F}_{ij}
\end{equation}
where $\vec{p}_i=m_i\vec{v}_i$, $ \vec{v}_i = d \vec{r}_i/dt$, and $\vec{r}_i$ denote the momentum, velocity, and position vectors of $i^{th}$ bead. The force $\vec{f}_i(t)$ on $i^{th}$ particle by all other $j$ particles consist of three pairwise additive components: (i) $ \vec{F}_{ij}^C$ represents the conservative force, (ii) $\vec{F}_{ij}^D$ denotes the dissipative force, and (iii) $\vec{F}_{ij}^R$ imply the random force such that $\vec{F}_{ij}=\vec{F}_{ij}^C+\vec{F}_{ij}^D+\vec{F}_{ij}^R$.

The conservative force describes the pairwise interaction between the beads. The most common and simple choice for $\vec{F}^C_{ij}$ is\cite{groot1997dissipative}
\begin{equation}
\label{cforce}
\vec{F}_{ij}^C=
\begin{cases}
a_{ij}\left(1-{r_{ij}}/{r_c}  \right)\hat{r}_{ij}, & r_{ij} < r_c \\
0, & r_{ij} \geq r_c
\end{cases} 
\end{equation}
which is a linear soft repulsive interaction up to a cut-off distance $r_c$. Here, $a_{ij}$ denotes the maximum repulsion of the conservative force, $\vec{r}_{ij}= \vec{r}_{j} - \vec{r}_{i}$, $r_{ij}= \lvert \vec{r}_{ij} \rvert $ and $\hat{r}_{ij}=\vec{r}_{ij}/r_{ij}$. The dissipative force $\vec{F}_{ij}^D $  has the following mathematical form
\begin{equation}
\label{dforce}
 \vec{F}_{ij}^D=-\gamma \omega_D(r_{ij})(\hat{r}_{ij} \cdot \vec{v}_{ij})\hat{r}_{ij} 
\end{equation}
where $\gamma$ is the friction coefficient or strength of viscus dissipation, $\omega_D(r_{ij})$ is the weight function of the dissipative force, and $\vec{v}_{ij}= \vec{v}_i-\vec{v}_j $. The effect of thermal fluctuation is described by\cite{groot1997dissipative} 
\begin{equation}
\label{rforce}
\vec{F}_{ij}^R=\sigma\omega_R(r_{ij})\sqrt{\Delta t}\xi_{ij} \hat{r}_{ij}
\end{equation}
where $\sigma$ is the strength of the random force, $\omega_R(r_{ij})$ is the corresponding weight function, and $\xi_{ij}$ denotes the Gaussian random variable with zero mean $\left\langle \xi_{ij}(t) \right\rangle=0 $ and unit variance $\left\langle \xi_{ij}(t)\xi_{kl}(t') \right\rangle=(\delta_{ik}\delta_{jl}+\delta_{il}\delta_{jk})\delta(t-t')$.\cite{groot1997dissipative, espanol1995statistical}. The symmetry relations $\vec{F}_{ij} = -\vec{F}_{ji}$, $\xi_{ij}=\xi_{ji}$, and the force $\vec{F}_{ij}^D$ and $\vec{F}_{ij}^R$ act along the line joining bead centres ensure the momentum conservation.\cite{groot1997dissipative, espanol1995statistical}.

To attain the correct canonical thermodynamic equilibrium state of a system at a temperature $T$, the strength of the random $(\sigma)$ and dissipative $(\gamma)$ forces must follow the fluctuation-dissipation relation\cite{espanol1995statistical,espanol2017perspective} $\sigma^2 = 2\gamma k_BT$ where $k_B$ is the Boltzmann constant; their corresponding weight functions must coupled as $\omega_D(r_{ij})=\left[\omega_R(r_{ij})\right] ^2$. The common choice for the weight function is\cite{groot1997dissipative} $
\omega_D(r_{ij})=\left(1-{r_{ij}}/{r_c}\right)^2$ for $r_{ij} < r_c $. However, we can choose other forms of weight functions given that the above two conditions are satisfied.

\subsection{Model parameters and other details}
\label{DPDmp}
The system evolution is followed by integrating the equations of motion shown in Eq. (\ref{Newton}) using a modified velocity-verlet algorithm.\cite{plimpton1995fast} In DPD simulations, cut-off radius $r_c$ is commonly considered as a characteristic length scale, and $k_BT$ as a characteristic energy scale.\cite{groot1997dissipative} Each DPD bead is considered to have equal mass $m_i = m$. The simulation results are presented in reduced DPD units with the parameters $r_c$, $m$, and $k_BT$, all set to $1.0$.\cite{groot1997dissipative} Following the dimensional analysis, the characteristic time scale is defined as $\tau=\left(mr_c^2/k_BT\right)^{1/2}$, which gives $\tau = 1.0$. We chose the integration time step $\Delta t=0.02$.\cite{singh2018role} The friction parameter value is set to $\gamma=4.5$, which is suitable for the system to reach the equilibrium temperature more rapidly and provide numerical stability for the specific time steps.\cite{groot1997dissipative} 

We consider the total number density of beads $\rho = 3$ appropriate for the DPD simulation of liquids; this choice of $\rho$ also ensures that the system is far away from the gas-liquid transition.\cite{groot1997dissipative, espanol1995statistical} The characteristic length and time scale in the actual units are estimated to be $r_c = 0.97 nm$ and $\tau = 8.3 ps$, respectively.\cite{singh2018role, singh2021photo} We set the value of repulsive interaction strength $a_{ij} = 25$ for the interaction between chemically compatible BCP beads (\textit{i.e.}, $a_{AA} = a_{BB}=25$); this choice is made to reproduce the compressibility of water by coarse-graining ten water molecules into one bead.\cite{groot1997dissipative, espanol1995statistical,espanol2017perspective, groot2006local} The interaction parameter, $a_{ij}=60$ for chemically incompatible beads (\textit{i.e.}, $a_{AB}=60$). The interaction of $A$ and $B$ beads with the wall beads is set to $a_{ij}=45$. In our DPD simulation, we quench the system at reduced temperature $T=1$, which we see in a short while that the chosen temperature is well below the critical temperature for the corresponding phase separation to take place.\cite{singh2018role, singh2021photo} 

We simulate BCP chains as bead-spring model where the beads of each sub-chains are connected by harmonic bond potential\cite{junghans2008transport,hsu2016static}, $E_b=\frac{k_b}{2}(r - r_0)^2$ where $k_b=128$ is the elastic constant, and $r_0=0.5$ is the equilibrium bond length.\cite{singh2021photo,nikunen2007reptational} The stiffness of the polymer chain is handled by the angle potential, $E_a = \frac{k_a}{2}(\cos\theta - \cos\theta_0)^2$, where $k_a=5$ is the strength of the angle potential. The angle between two successive bonds along the chain is $\theta$. We have taken the equilibrium value $\theta_0=180$.\cite{nikunen2003would,nikunen2007reptational,junghans2008transport,hsu2016static,plimpton1995fast}
Since DPD utilizes softcore interaction potential between the beads, there could be an unphysical bond crossing between the BCP chains. To mitigate that, we used \textit{modified segmental repulsion potential} (mSRP) \cite{sirk2012enhanced}. In this formulation, pseudo beads are considered at the center of all bonds. These beads interact with a softcore repulsive interaction\cite{sirk2012enhanced}, $\vec{F}_{ij}^{mS} = k_{mS}\left(1-{r_{ij}}/{r_c^{mS}}  \right)\hat{r}_{ij}$, for $r_{ij} < r_c^{mS}$ where $k_{mS}= 100$ (in reduced DPD units) is the force constant. The cutoff distance for the mSRP interaction is taken as $r_c^{mS}=0.6$. We use the LAMMPS simulation package\cite{plimpton1995fast} with mSRP code\cite{sirk2012enhanced} to integrate the equations of motion.

The box size used in our DPD simulation is $L_x\times L_y\times L_ z= 64 \times 64 \times 64 $; thus, the total number of DPD beads in the simulation box is $N=\rho\times L^3 = 7,86,432$. The period boundary conditions are applied in $x$- and $y$- directions, whereas two solid walls bound the simulation box in the transverse $z$-direction. The height of each wall is set to $h = 1.0$. The walls are also made of DPD beads with a number density $\rho_w = 3$; therefore, the total number of wall beads is $N_w=24,576$, so the volume fraction is $\phi_w = 3.125\times 10^{-2}$. The bounce-back boundary condition is applied at the fluid-wall interface\cite{esteves2013surface} to prevent the penetration of fluid beads into the walls.\cite{liu2015designing} For the simulation, we consider a critical BCP melt $(n:m=1:1)$ with chain length $N_p= 32$ (degree of polymerization). Total number of $A$- and $B$-type beads is $N_A = N_B = 3,80,928$ ($\phi_A = \phi_B = 4.843\times 10^{-1}$), \textit{i.e.},  $\rho_A = \rho_B = 1.453$. The total number of BCP chains in the simulation box is $N_{BCP} = 23,808$. 

To shear the phase separating BCP melt, we let the walls move in $x$-direction with a constant velocity, $\vec{v}_w \in \left(v_{wx}, v_{wy}=0, v_{wz}=0 \right)$. When walls are fixed, corresponding velocity components $v_{wx}$, $v_{wy}$, and $v_{wy}$ are set to zero (case 1). We consider three other cases depending on the direction of the wall's motion: (i) top wall moves in the positive $x$-direction (case 2) with $v_{wx} = 0.1$, $0.5$, and $1.0$, while the bottom wall is fixed. (ii) The top and the bottom walls move in the same direction (case 3) with the same wall velocities as in case 2. (iii) Both walls move in the opposite directions, the top wall moves in the positive $x$-direction, and bottom wall in the negative $x$-direction; we set this condition as case 4. 

To demonstrate the effect of shear, we compute shear viscosity, $\eta$ using the technique developed by Muller-Plathe \cite{muller1999reversing,kelkar2007prediction}. In this technique, momentum flux is imposed on the system by the exchange of momentum of atoms along a particular direction (say $z$-direction). The velocity profile is generated as a response to momentum flux. Shear viscosity can be calculated from the slope of velocity profile and imposed momentum flux using the equation: $j_z(p_x) = -\eta \left({\partial v_x}/{\partial z}\right)$, here $j_z(p_x)$ is the momentum flux, and ${\partial v_x}/{\partial z}$ is the velocity gradient or shear rate. Further, we quantitatively evaluate the anisotropic degree of evolved morphology by calculating the anisotropy parameter $D_{xy}(t)$ as \cite{Onuki86, zhang2022microstructural}; 
\begin{equation}
	\label{anisotropy}
	D_{xy}(t) = \frac{\sum (k_x^2 - k_y^2)/k^2) S(k_x, k_y, k_z)}{\sum S(k_x, k_y, k_z)}. 
\end{equation}
Here, $k_x$ and $k_y$ are the components of wave vector $\vec{k}$ in the $xy$-plane for a fixed $k_z$ where $k^2 = k_x^2 +k_y^2+k_z^2$, and $S(k_x, k_y, k_z)$ denotes the structure factor at a point $\left( k_x, k_y, k_z\right)$ in Fourier space. Similarly, we calculate the anisotropy parameter, $D_{yz}(t)$, and $D_{xz}(t)$ in the $yz$ and  $xz$ planes, respectively. Note that, for a perfect isotropic system, $D_{ij}\simeq 0$ and for a fully anisotropic system, $D_{ij}\rightarrow 1$.

\subsection{Morphology characterization and scaling functions}
\label{scaling}
The two most desirable and experimentally relevant functions to characterize the evolution morphology are the correlation function and its Fourier transform, the structure factor.\cite{bray2002theory,puri2009kinetics} We introduce the two-point equal-time correlation function as
\begin{equation}
\label{corr}
C(\vec{r},t)= \left\langle \psi\left(\vec{r}_1,t\right)\psi\left(\vec{r}_2,t\right)\right\rangle-\left\langle \psi\left(\vec{r}_1,t\right)\right\rangle \left\langle \psi\left(\vec{r}_2,t\right)\right\rangle
\end{equation}
where $\psi(\vec{r},t)$ denotes the order parameter. The structure factor is defined as 
\begin{equation}
\label{stru}
S(\vec{k},t)=\int d\vec{r}\exp\left(i\vec{k}\cdot\vec{r}\right)C(\vec{r},t)
\end{equation}
where $\vec{k}$ is the scattering wave vector. 

The order parameter $\psi(\vec{r},t)$ is obtained as the local density difference of the constituents located at $\vec{r}$ at a given time $t$\cite{singh2015phase,singh2014kinetics},
\begin{equation}
\label{op}
\psi(\vec{r},t)=\frac{n_A(\vec{r},t)-n_B(\vec{r},t)}{n_A(\vec{r},t)+n_B(\vec{r},t)}
\end{equation} 
where $n_A(\vec{r},t)$ and $n_B(\vec{r},t)$ represent the local number of $A$ and $B$ types of beads calculated as follows: we divide the simulation box into non-overlapping unit cubes and then count the number of $A$ and $B$ types of beads in each unit cube. Thus, from Eq. (\ref{op}) $\psi(\vec{r},t)\in (0, 1)$ for $n_A(\vec{r},t)> n_B(\vec{r},t)$, and $\psi(\vec{r},t)\in (0, -1)$ for $n_A(\vec{r},t)< n_B(\vec{r},t)$. However, for $n_A(\vec{r},t)= n_B(\vec{r},t)$, we assign $\psi(\vec{r},t)\in (0, 1)$ or $\psi(\vec{r},t)\in (0, -1)$ with equal probability \cite{singh2015phase}. 

The number of peaks in the structure factor or the oscillatory behavior of the correlation function after zero-crossing signifies the periodicity of evolving morphology in the phase separating BCP melts. In our simulation, we spherically average $C(\vec{r},t)$ and $S(\vec{k},t)$ for better statistics. The corresponding quantities are denoted by $C(r,t)$ and $S(k,t)$, respectively. The characteristic length scale, $R(t)$ is defined as the distance over which $C(r,t)$ decays to a fraction of its maximum value ($C(r,t)\vert_{r=0} =1$). There are a few other definitions of $R(t)$, but they are equivalent in the asymptotic limit; they differ only by some constant multiplicative factors in the scaling regime.\cite{oono1987computationally,singh2011control} Herein, the distance over which the correlation function decays to $0.2$ provides a good measure of $R(t)$.\cite{oono1987computationally,oono1988study} 

The basic assumption of scaling is the existence of a unique characteristic length scale $R(t)$. The dynamical scaling form of the correlation function and the structure factor has the following form\cite{bray2002theory,puri2009kinetics}:
\begin{equation}
\label{scorr}
C(r,t) = f(r/R(t))
\end{equation}
\begin{equation}
\label{sstru}
S(k,t)=R(t)^d g(kR(t))
\end{equation}
where $f(r/R(t))$ and $g(kR(t))$ are the corresponding scaling functions. The characteristic length scale follows the power-law growth: $R(t)=R_0+at^\phi$ where $R_0$ is the average domain size in the transient growth regime, which is taken as an offset value, therefore, we define $R(t)\equiv R(t) - R_o$\cite{singh2018role}, and $a$ is a growth rate constant. To find the asymptotic growth exponent ($\phi$), we plot $R(t)$ versus $t$ on a logarithmic scale and estimate the slope of the curve. Further, we follow another intuitive way to verify the same by computing the effective growth exponent as
\begin{equation}
\label{gexp} 
\phi_{eff} = \log_{\alpha}\left[R\left(\alpha t \right)/ R\left(t \right)\right]
\end{equation} 
and plot it against inverse of domain size $1/R(t)$ where we set the log-base $\alpha = 2$.\cite{huse1985pinning}

\section{Numerical results}
\label{Nr}
This section will discuss the numerical results obtained for microphase separating symmetric BCP melt ($A_nB_n; n=16$) confined within two parallel solid walls in the transverse $z$-direction. At the onset of the DPD simulation, we first equilibrate the initial system for $t=2\times 10^5$ at a high-temperature $T=5$ (in reduced DPD units) to prepare a homogeneous mixture. We then reset the time at $t=0$, quench the system at a lower temperature $T=1$, and monitor the evolution pattern at different times. 

\subsection{Both walls are fixed}
\label{bwf}
\begin{figure}[tb]
	\centering
	\includegraphics[width=0.7\textwidth]{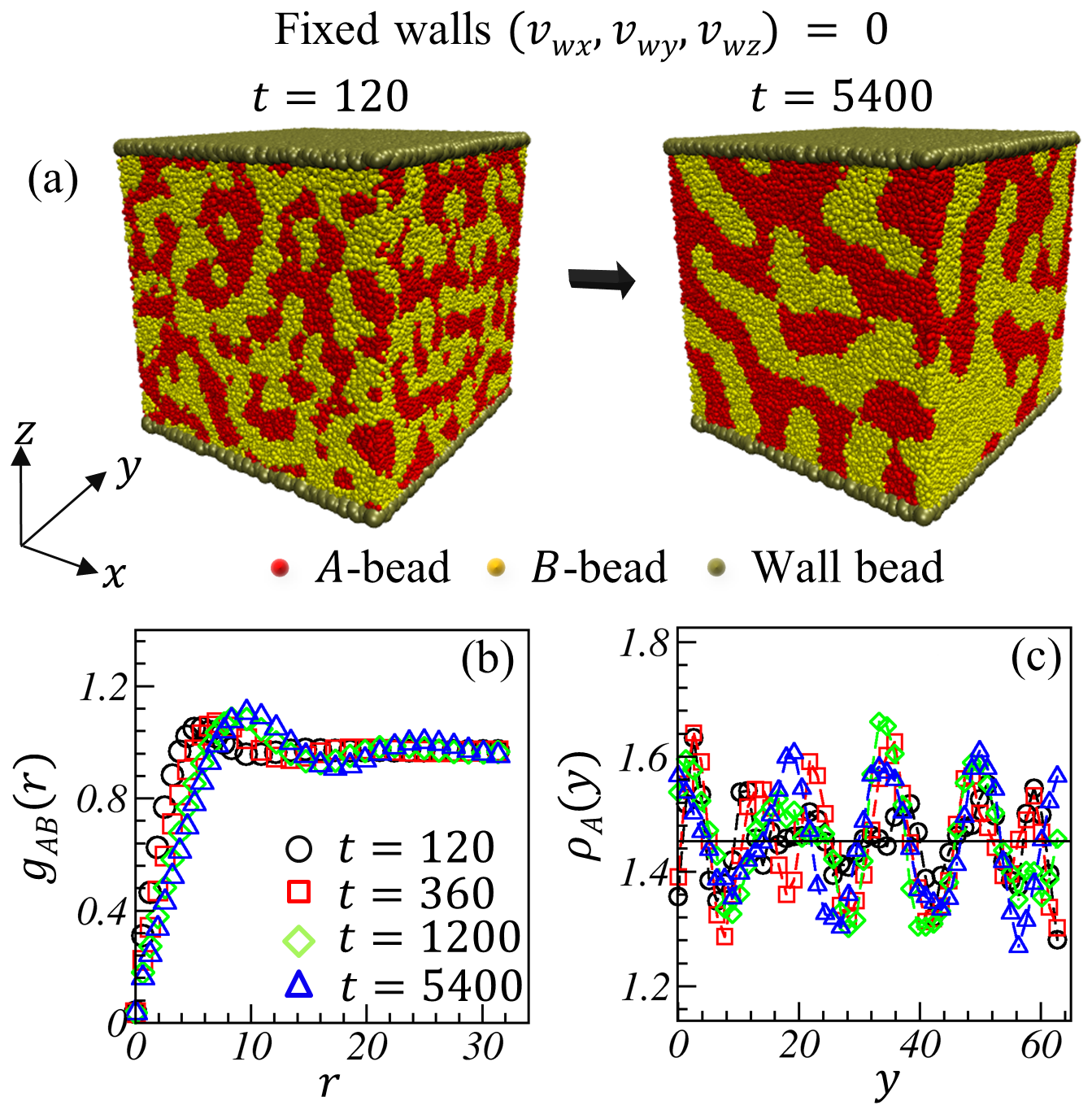}
	\caption{The system is confined within fixed walls at the top and bottom, as indicated by the olive beads. (a) Evolution morphology of BCP melt at $ t=120$ and $t=5400$. (b) Comparison of radial distribution function, $g_{AB}(r)$ versus $r$ at different evolution times. (c) Average number density distribution, $\rho_A(y)$ of $A$-beads along $y$-direction for the same time steps as in (b).}
	\label{fig1}
\end{figure}

To begin with, we first study case 1 with both walls fixed (no shear) to put the rest of the discussions in proper context. The evolution morphology at $t=120$ and $t=5400$ are shown in Fig. \ref{fig1}(a); $A$-type beads are in red, $B$-type beads are in yellow, and the olive color marks the wall beads. We plot the corresponding radial distribution function (RDF), $g_{AB}(r)$ vs. radial distance, $r$ in Fig. \ref{fig1}(b) at different time steps as denoted by the various symbols. The extension in peak height and the shift to higher $r$ with time indicate the growth of $A$ and $B$ clusters. The black curve ($t = 120$) shows a single lower peak at a small $r$, which saturates at larger $r$ values; this demonstrates an early-time evolution of homogeneous BCP melt into tiny domains. As time progresses, $g_{AB}(r)$ begins to develop secondary peaks as displayed by the red ($t = 360$), green ($t=1200$), and blue ($t = 5400$) curves, respectively; this indicates the coarsening of BCPs into periodic structures. Further, the local number density profile of $A$-type beads, $\rho_A(y)$ along the  $y$-direction is displayed in Fig. \ref{fig1}(c) for the same time sets as in Fig. \ref{fig1}(b). The solid black line shows the actual number density of $A$ beads ($\rho_A = 1.453$). The increasing density profile amplitude demonstrates the growth of phase separating domains, and the oscillatory variation ascertains the periodic structure formation. The density profiles in other directions are not shown here due to brevity.

\begin{figure}[ht!]
	\centering
	\includegraphics[width=0.7\textwidth]{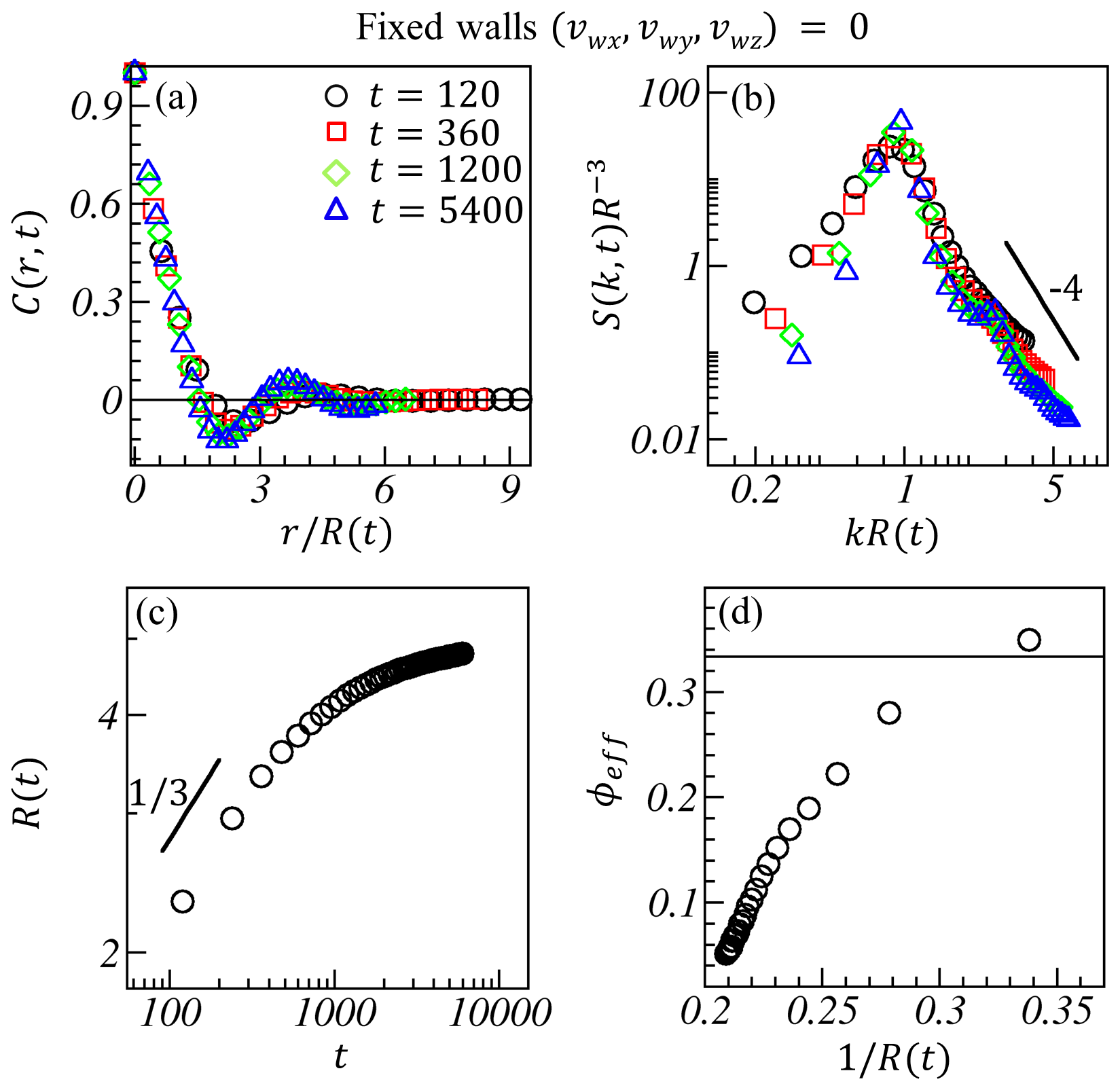}
	\caption{Spherically averaged scaled functions and length scale for the evolution displayed in Fig. \ref{fig1}. (a) $C(r,t)$ versus $r/R(t)$ at different times for the fixed walls. The solid line is a guide for the zero-crossing of $C(r,t)$. (b)  $S(k,t)$ versus $kR(t)$ on a logarithmic scale for the same times as in (a). The solid line with slope $-4$ represents the Porod's law: $ S(k,t)\sim k^{-4}$ at $ k\rightarrow \infty$. (c) Characteristic length scale, $R(t)$ versus $t$ for the same data sets as in (a) and (b). The solid line shows the diffusive growth law $(\phi\sim1/3)$ at early times, which gradually saturates later, illustrating usual microphase separation kinetics in BCP melt. (d) Presents effective growth exponent $\phi_{eff}$ versus $1/R(t)$. The solid line shows the reference value of $\phi_{eff}\sim 1/3$ for early diffusive growth.}  
	\label{fig2}
\end{figure}

We plot the scaled correlation functions, $C(r, t)$ vs. $r/R(t)$ in Fig. \ref{fig2}(a) at four different times. The correlation functions at early times, $t = 120$ (black symbol) and $t = 360$ (red symbol), deviate from the scaling. Note that early-time domain evolution in BCP melt is analogous to spinodal decomposition in polymer blends.\cite{groot1999role} An excellent data overlap of the correlation functions at late times, $t = 1200$ and $5400$ (denoted by the green and blue symbols), justify the dynamical scaling regimes. A more oscillatory form of $C(r, t)$ after the zero crossing characterizes the development of periodic structures in the late times. The solid black line denotes the zero-crossing of $C(r, t)$. Notice that we do not observe a clear lamellar morphology within the given simulation period as displayed in Fig. \ref{fig1}(a) for the symmetric BCP melt with chain length $N_p= 32$ ($A_{16}B_{16}$). Since diffusion coefficient, $D\sim 1/N_p$,  the formation of lamellar morphology is observed for symmetric BCP melts with shorter chains within the same simulation period, as displayed in Fig. \ref{SI1}(a) for $N_p=8$ ($A_{4}B_{4}$) and in Fig. \ref{SI1}(b) for $N_p=16$ ($A_{8}B_{8}$). Therefore, we need longer steps to observe lamellar morphology for BCPs with chain length $N_p= 32$. The scaled structure factor, $S(k,t)R(t)^{-3}$ vs. $kR(t)$ in Fig. \ref{fig2}(b) is displayed for the same times as in Fig. \ref{fig2}(a). As expected, we observe a crossover in the scaling functions. The structure factor data deviates at early times (indicated by the black and red symbols). Notice the narrowing peak in $S(k,t)$ and the appearance of the shoulder (at $kR(t)\simeq 4$) in the green and blue curves at $t = 1200$ and $t =5400$; this verifies the formation of periodic morphology at the late stages of coarsening. The structure factor shows a power-law decay, $S(k, t)\sim k^{-(d+1)}$ at large wave vector ($k\rightarrow\infty$) regimes where $d$ is the system's dimension; this is known as Porod's law\cite{porod1982small} results from scattering off sharp interfaces.

Next, to study the domain growth law for the evolution shown in Fig. \ref{fig1}(a), we plot the characteristic length scale, $R(t)$ vs. $t$ in Fig. \ref{fig2}(c), on a log-log scale. BCP melt follows the diffusive growth (i.e., $\phi \sim 1/3$) in an early spinodal time window up to $t_{sp} \simeq 800$ such that $R(t_{sp}) \sim 4$ (equivalent to the average size of BCP sub-chains). The length scale then crosses over to saturation when the constraints imposed by the topological connection become evident. The saturating $R(t)$ marks the freezing microphase morphology for symmetric BCPs. A more precise way of computing the growth exponent is to plot $\phi_{eff}$ vs. $R(t)^{-1}$ from Eq. (\ref{gexp}) and extract $\phi_{eff}$ from data extrapolation in the asymptotic limit.  Such an exercise is shown in Fig. \ref{fig2}(d), which demonstrate that in the limit $R(t)^{-1} \rightarrow 0$, the growth exponent converges to $\phi_{eff} \rightarrow 0$.

\subsection{Top wall moves}
\label{twm}
Let us return to the phase separation of BCP melt under the influence of shear arising due to moving walls. First, we focus on case 2, where the top wall moves in the positive $x$-direction, and the bottom wall is fixed. Any variation in the wall velocity can alter the shear rate, $\dot{\gamma}\propto v_{wx}$; here we choose three different wall velocities, $v_{wx} = 0.1$, $0.5$, and $1.0$.

The evolution morphology for case 2 is displayed in Figs. \ref{SI2}. It is apparent that with the increase in $v_{wx}$, domains of $A$ and $B$ sub-chains adjacent to the top wall appear to flow more along the shear direction. Recall that in case 1, the system could not develop lamellar morphology up to simulation time $t = 5400$. Herein, lamellar slabs appear earlier with increasing shear rates; the reason could be the flow and orientation of microdomains in the shear direction. Therefore, due to the competition of unidirectional shear and simultaneous microphase separation kinetics, BCP chains segregate perpendicular to the shear direction to form a more stable lamellar morphology than in any other adjustments. In other words, by adjusting normal to the shear direction, $A$ and $B$ domains try to overcome the effect of shear. Since only the top wall moves, the low shear rate ($v_{wx} = 0.1$) primarily influences a small upper region of the simulation box. Thus, lamellar formation begins closer to the moving wall along the flow direction and perpendicular to the $yz$-plane. A few more layers in the simulation box are affected by the increase in wall velocity, $v_{wx} = 0.5$ and $v_{wx} = 1.0$, and therefore, we get more ordered lamellar strips within the same time interval. The unidirectional shear enhances the tendency of lamellar pattern formation for a phase separating symmetric BCP melt.

\begin{figure}[ht!]
	\centering
	\includegraphics[width=0.7\textwidth]{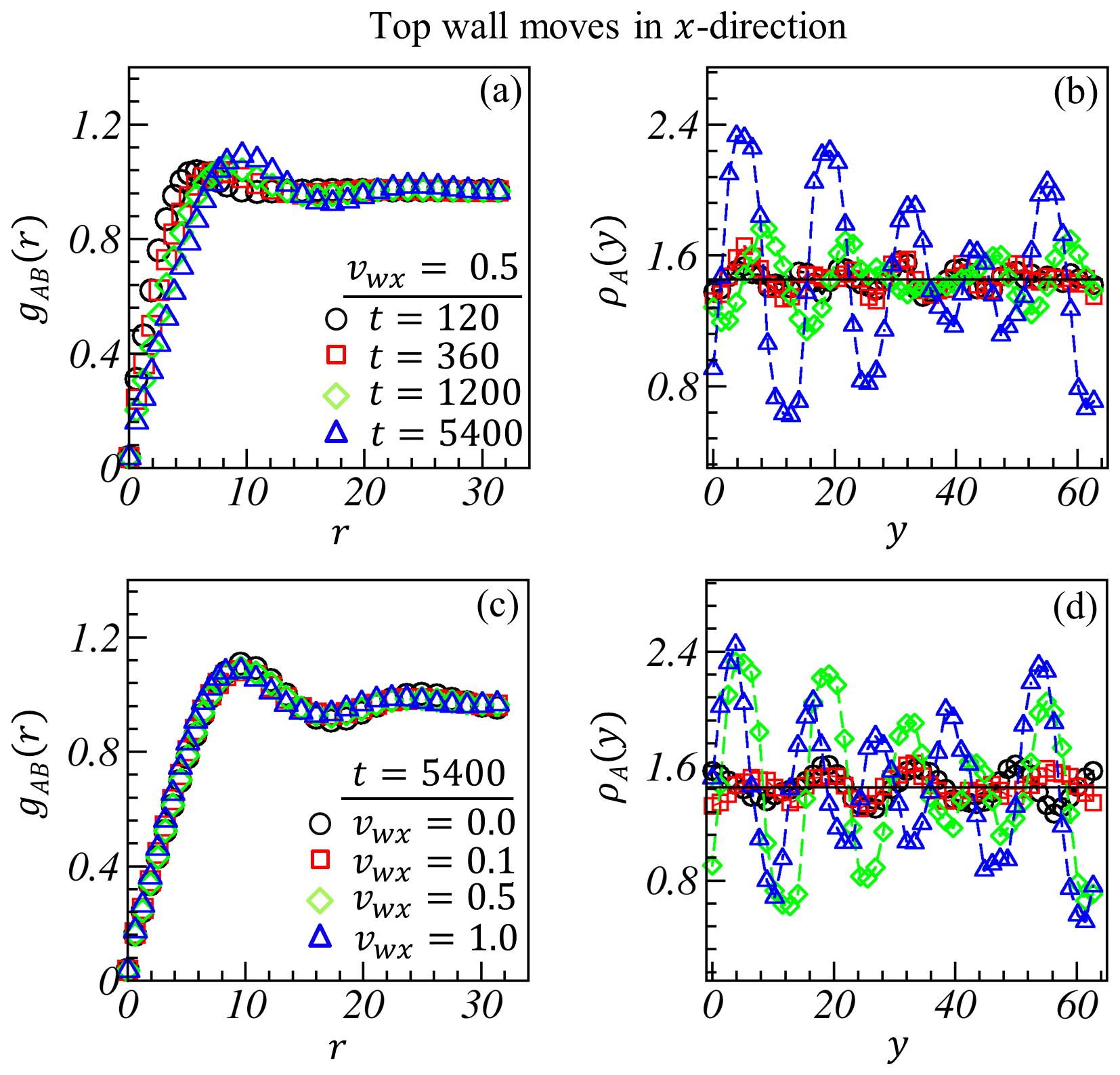}
	\caption{Top wall is moving with velocity $v_{wx} =0.5$. (a) Plots for $g_{AB}(r)$ versus $r$ at different times, indicated by various symbols. (b) $\rho(y)$ of $A$-beads along $y$-direction for the same time steps as in (a). (c) $g_{AB}(r)$ versus $r$ for different top wall velocities for case 2 at a fixed $t=5400$. (d) $\rho(y)$ versus $y$ for the same set of data as in (c).}
	\label{fig3}
\end{figure}

To characterize the evolution morphology illustrated in Fig. \ref{SI2}, we plot $g_{AB}(r)$ vs. $r$ in Fig. \ref{fig3}(a) at different time steps for $v_{wx}=0.5$. The black curve showed early time data when a few small microdomains evolved via spinodal decomposition. Thus, a single peak is observed at a shorter distance $r \simeq 5.0$. The main peak strength of $g_{AB}(r)$ increases with time and shows a relatively higher amplitude at $r \simeq 9.25$ at the late time (see the blue curve) than any other curves at earlier times. A secondary peak in $g_{AB}(r)$ is also clearly visible in the blue curve at a distance $r\simeq 24.0$. The main reason is the coarsening of lamellar morphology. Corresponding local number density profile, $\rho_A(y)$ is plotted against the $y$-direction in Fig. \ref{fig3}(b). A much lower density profile amplitude at early times verifies tiny microdomain formation (see the corresponding black and red curves). However, at late times, a more considerable amplitude and oscillatory behavior of $\rho_A(y)$ are observed due to larger and periodic domain coarsening.

Further, we compare $g_{AB}(r)$ vs. $r$ in Fig. \ref{fig3}(c) to characterize the evolved morphologies at a late time for different wall velocities. All the RDF curves present an excellent data overlap closer to the first peak; however, the main peak width slightly narrows as its position shifts to smaller $r \simeq 9.6$, $9.6$, $9.25$, and $9.0$ with increasing shear rates. Subsequently, the secondary RDF peaks, which are developed due to the formation of periodic morphology, also shifted to its left at $r \simeq 25.0$, $24.32$, $24.0$, and $22.4$ for $v_{wx}=0.0$, $0.1$, $0.5$, and $1.0$, respectively. Overall, these plots suggest increasing periodicity and the thinning of domain sizes with increasing shear rates. The much lower density profile ($\rho_A(y)$) amplitude at smaller shear rates ($v_{wx}=0.0$, $0.1$), indicated by the back and red symbols in Fig. \ref{fig3}(d), imply that the periodic structures are not a fully developed lamellar$-$many short and randomly oriented domain stripes are still present in the system. In contrast, $\rho_A(y)$ gains significant height and periodicity at higher shear rates ($v_{wx}=0.5$, $1.0$),  suggesting the evolution of appreciable lamellar morphology.

\begin{figure}[ht!]
	\centering
	\includegraphics[width=0.7\textwidth]{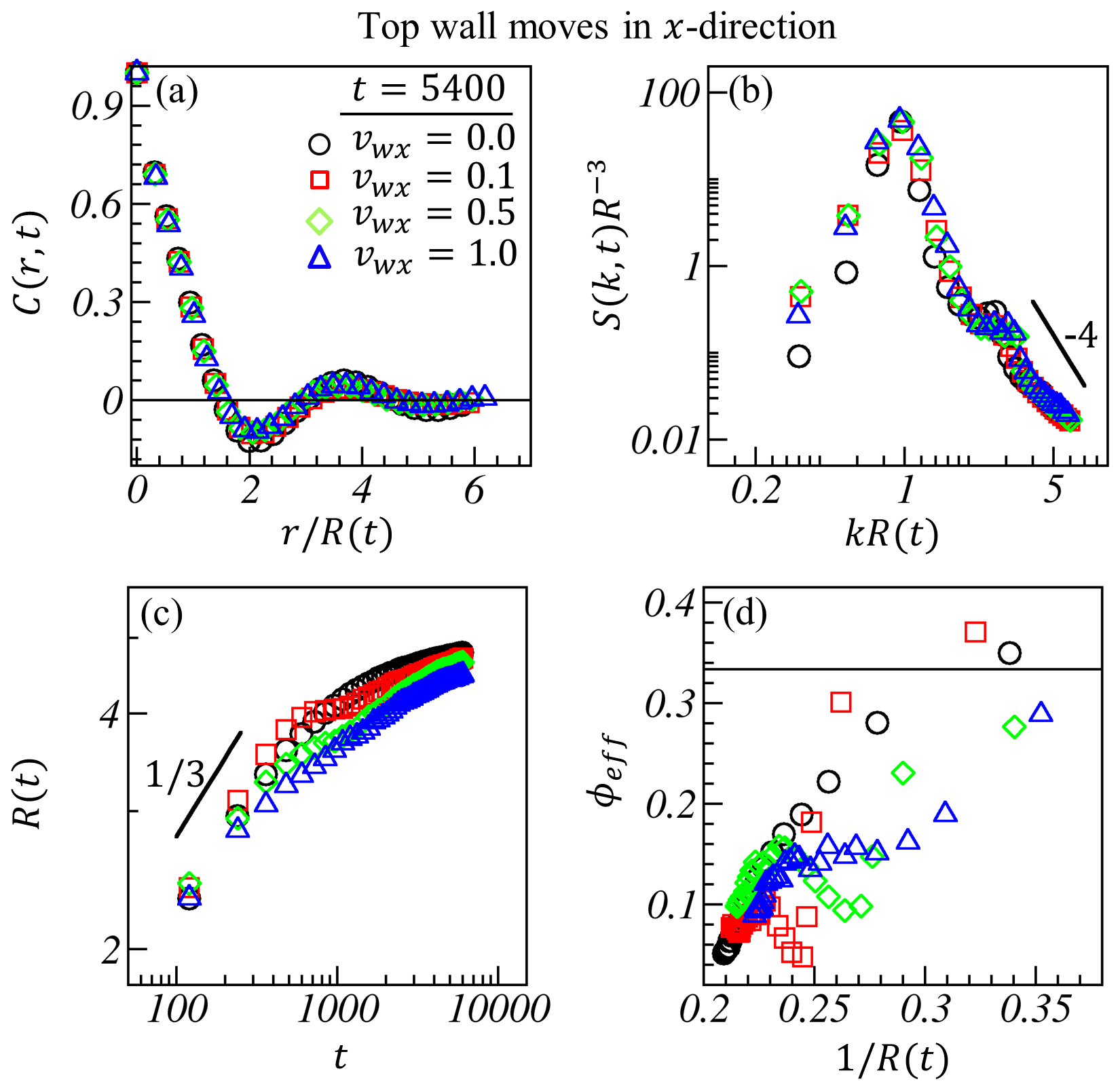}
	\caption{Comparing scaling function, length scale, and effective growth exponent at $t=5400$ for different wall velocities for case 2 as indicated by the various symbol types. $C(r,t)$ versus $r/R(t)$ is displayed in (a), and corresponding $S(k,t)$ versus $kR(t)$ on a logarithmic scale in (b). (c) $R(t)$ versus $t$ on a log-log scale at different wall velocities for the evolution shown in Fig. (\ref{SI2}). The variation of growth exponent $\phi_{eff}$ as a function of $1/R(t)$ is in (d).} 
	\label{fig4}
\end{figure}

To see the effect of various shear rates on the dynamic scaling functions for case 2, we plot $C(r,t)$ vs. $r/R(t)$ in Fig. \ref{fig4}(a) and $S(k,t)$ vs. $kR(t)$ on a log-log scale in Fig. \ref{fig4}(b) for different wall velocities at $t = 5400$. The oscillatory $C(r,t)$ curves in Fig. \ref{fig4}(a) characterize the developing periodic morphology at a late time; the appearance of the distinct shoulder at large $kR(t)$ for each curve in Fig. \ref{fig4}(b) verify the same. Notice the slight decrease in amplitude of oscillation in $C(r,t)$ and the broadening in $S(k,t)$s' main peak for $v_{wx}\neq 0.0$. Since the first moment of $S(k,t)$ is inversely proportional to average domain size ($\langle k \rangle \sim R(t)^{-1}$) \cite{bray2002theory, puri2009kinetics}, the broadening of main $S(k,t)$ peak implies an increase in $\langle k \rangle$ with shear rate justifies shear-thinning at late times. Further, the shoulder in $S(k,t)$ shows up at smaller $kR(t)\simeq 2.67$ for $v_{wx}=0.0$ (black curve), whereas it appears at larger $kR(t)\simeq 3.06$ in the blue curve. Overall, these observations indicate the shear-thinning of morphologies. We notice that the black curves in Figs. \ref{fig4}(a-b) slightly deviate from the scaling. However, for $v_{wx} \neq 0.0$, the scaling function data illustrates a good overlap; hence, they belong to the same universality class. A solid line with slope $-4$ in Fig. \ref{fig4}(b) shows that the scaled structure factor tail follows Porod's law for all cases.

Next, we compare the characteristic length scale, $R(t)$ vs. $t$ on a logarithmic scale in Fig. \ref{fig4}(c) and the corresponding effective growth exponent in Fig. \ref{fig4}(d) for $v_{wx} \neq 0.0$ values. The black curve for $v_{wx}=0.0$ is plotted for reference. Note that domain coarsening and their flow along the shear direction are two physical processes occurring simultaneously. The lower shear rate ($v_{wx}=0.1$, red curve) seems to complement the early-time domain growth ($R(t)\sim t^{1/3}$). Thus, we observe a larger average domain size up to the time, $t_{sp} \simeq 600$, indicating lower shear-thinning than other higher shear rates. The topological constraint becomes apparent for $t>t_{sp}$. We notice a dip in the growth scale for a certain period ($600<t<1000$) where $A$ and $B$ domains could be adjusted in a shear direction to minimize its effect. The length scale grows again and then crosses over to saturation, indicating the freezing in microdomain growth to a smaller length scale than in case 1 ($v_{wx}=0.0$), which marks the shear-thinning effect. The shear effect becomes prominent with increasing wall velocity, which causes domain thinning from the early times. Thus, the overall growth rate decreases. Early-time growth exponent reduces from a pure diffusive growth $R(t)\sim t^{1/3}$, then crosses over to frozen microphase morphology at the late stages. The effective growth exponent, $\phi_{eff}$ vs. $1/R(t)$ curves in Fig. \ref{fig4}(d) again illustrate the same; $\phi_{eff} \rightarrow 0$ in the limit $1/R(t) \rightarrow 0$.

\subsection{Both walls move in the same direction}
\label{bwm_sd}
Now, we turn our attention to case 3, where we shear the segregating BCP melt from both sides in the same direction and monitor its effect on the evolution morphology (displayed in Fig. \ref{SI3}) and the growth kinetics. The result of shear on the morphology is more prominent at $v_{wx}=0.5$ and $1.0$. The system begins to develop periodic structures from the top and bottom region of the simulation box, particularly closer to the walls where shear is more effective than in other areas. A more explicit lamellar morphology seemed to form herein than in case 2 at the same wall velocities. For lower velocity $v_{wx}=0.1$, the effect of shear on the evolving morphology emerges almost similar to what it was in case 2.
 
\begin{figure}[ht!]
	\centering
	\includegraphics[width=0.7\textwidth]{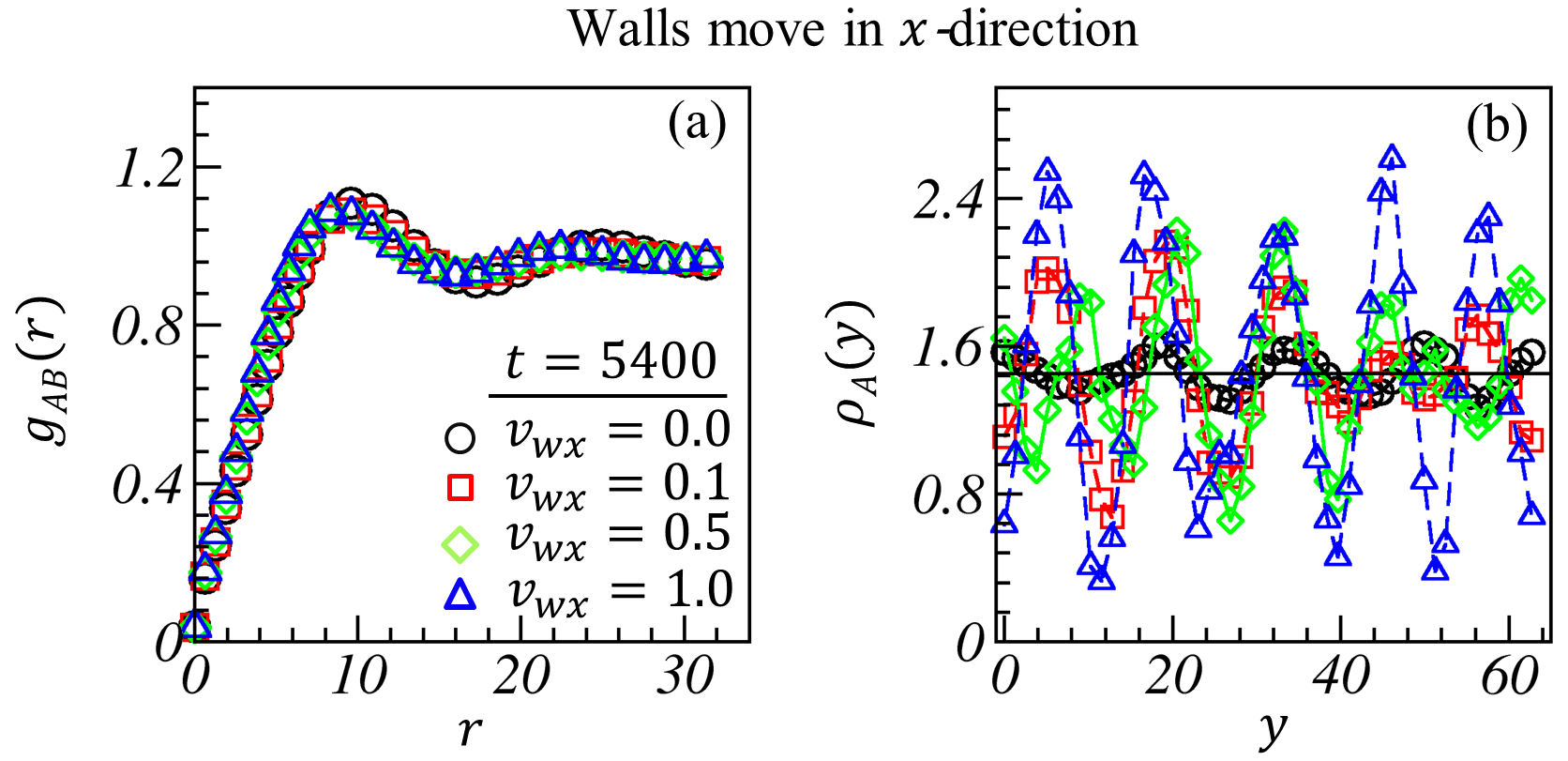}
	\caption{Case 3: Both walls are moving with the same velocities in $x$-direction. (a) $g_{AB}(r)$ versus $r$, and (b) average number density profiles of $A$-type beads in $y$-direction at $t=5400$ for different walls velocities are presented with various open symbols; related morphologies are shown in Fig. (\ref{SI3})}
	\label{fig5}
\end{figure}

We plot $g_{AB}(r)$ for different wall velocities in Fig. \ref{fig5}(a) at $t =5400$. Similar to the previous cases, $g_{AB}(r)$ becomes more oscillatory with increasing shear rate, signifying increasing periodicity. The peak positions of $g_{AB}(r)$ move to their left with increasing shear rates (denoted by various symbols); the primary peak positions are at $r \simeq 9.6$, $9.5$, $9.0$, $9.0$, and the secondary peaks are at $r \simeq 25.0$, $24.32$, $23.04$, $22.4$, respectively. This indicates the thinning in average domain size with shear rates. The local density variation in Fig. \ref{fig5}(b) illustrates the equal distribution of $\rho_A(y)$ around its average value ($\rho_A = 1.453$). Their enhanced amplitude than previous cases signifies the formation of a more well-formed lamellar morphology, perpendicular to the $y$-axis; see the green and blue curves at $v_{wx}=0.5$ and $v_{wx}=1.0$, respectively. Notice the red curve at $v_{wx}=0.1$, showing an apparent enhancement in  $\rho_A(y)$ than in case 2; thus, we get a more ordered structure in case 3 than in case 2 with the same shear rates.

\begin{figure}[ht!]
	\centering
	\includegraphics[width=0.7\textwidth]{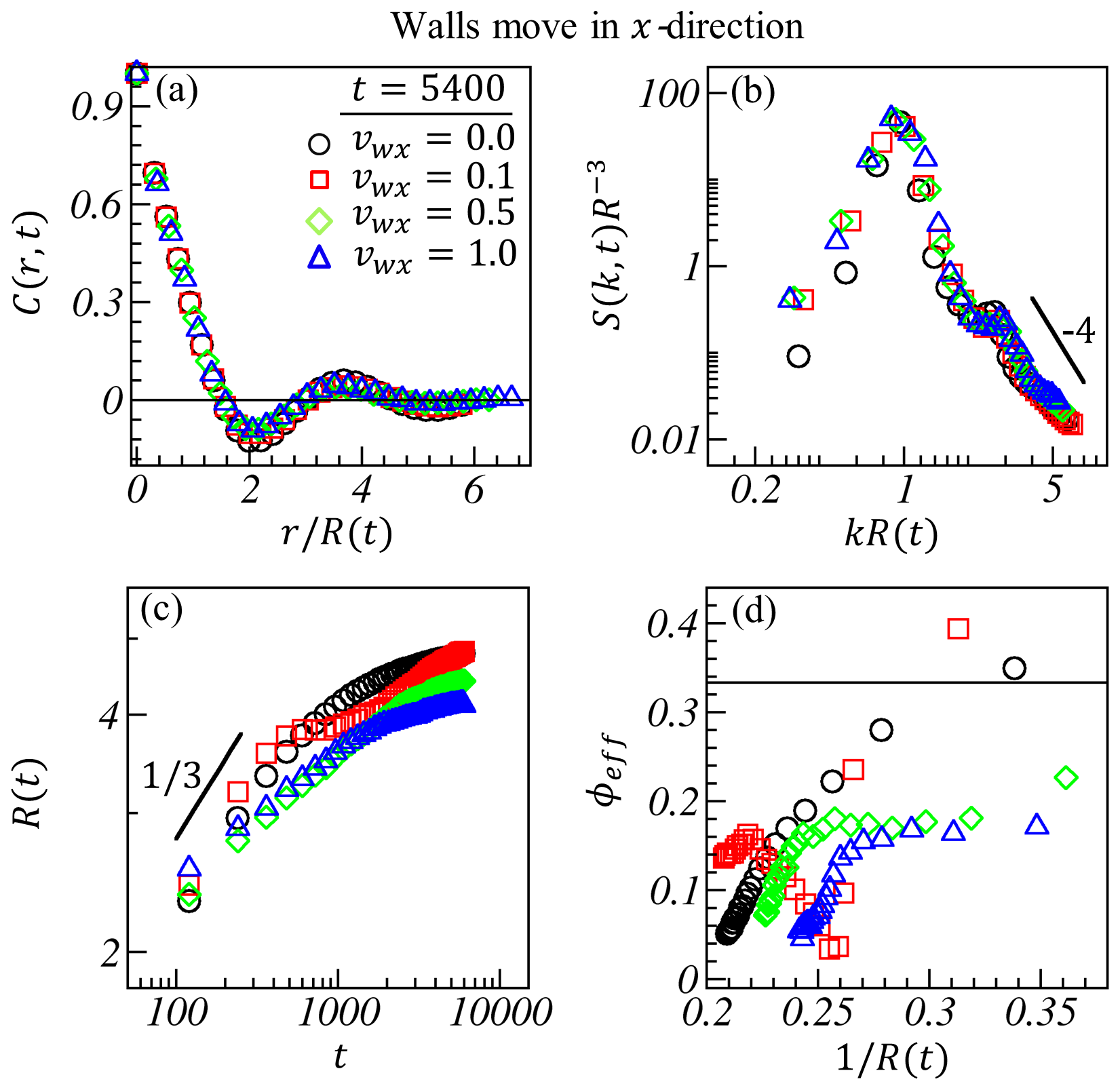}
	\caption{(a) Comparison of $C(r,t)$ versus $r/R(t)$ at different wall velocities for the morphologies illustrated in Fig. (\ref{SI3}). Corresponding $S(k,t)$ versus $kR(t)$ curves are in (b) on a logarithmic scale. The average domain size, $R(t)$ versus $t$, and related effective growth exponents  $\phi_{eff}$ versus $1/R(t)$ are displayed in (c) and (d), respectively.}
	\label{fig6}
\end{figure}

The late time dynamic scaling functions, $C(r, t)$ and $S(k, t)$ in Fig. \ref{fig6}(a) and Fig. \ref{fig6}(b) exhibit apparent deviation at $v_{wx}=0.1$ (red  symbol) from the reference case (black curve) including the high shear cases, as displayed in green and blue symbols. However, $C(r, t)$ and $S(k, t)$ curves at higher shear rates illustrate excellent data overlap, thus, exhibiting dynamical scaling. The structure factor shows Porod's tail ($S(k,t) \sim k^{-4}$) for all the shear rates for $kR(t) \rightarrow \infty$, indicating the sharp domain interface between $A$ and $B$ phases. The oscillation in the correlation function and the emergence of the shoulder at large $kR(t)$ in the structure factor represent the formation of the periodic structures. After the zero crossing, the smaller amplitude of oscillation in $C(r, t)$ and the broadening of the prominent peak in $S(k, t)$ with shear rates attribute to the shear-thinning of evolving morphologies.
    
Similar to case 2, the low shear rate complements the early time diffusive growth, and we observed a slightly larger average domain size up to $t_{sp} \simeq 600$ (see the red curve in Fig. \ref{fig6}(c)). Then, a flattening of length scale data (\textit{i.e.}, negligible domain growth) is observed up to $t \simeq 1000$ due to the competing process of shear and phase separation kinetics. Corresponding growth exponent ($\phi_{eff}$) shows a significant dip (see the red curve in Fig. \ref{fig6}(d)). The BCP sub-chains are most likely arranged normal to the flow direction in this period to minimize the shear effect. Beyond this, domains begin to grow again with a lower growth rate due to shear-thinning than the reference case during the same period. The length scale tends to saturate at a slightly larger value than the reference case (see the red curves in Figs. \ref{fig6}(c-d)). At higher wall velocities, the shear effect is prominent from the beginning; the growth exponent ($\phi_{eff}\simeq 0.2$) is smaller, as displayed by the green and blue curves in Figs. \ref{fig6}(c-d). The length scale tends to saturate to the lower values at late times, justifying the shear-thinning. Again, the shear-thinning is more prominent for the higher shear rate. Thus, in the limit of $R(t)^{-1} \rightarrow 0$, the effective growth exponent $\phi_{eff} \rightarrow 0$.

\begin{figure}[ht!]
	\centering
	\includegraphics[width=0.7\textwidth]{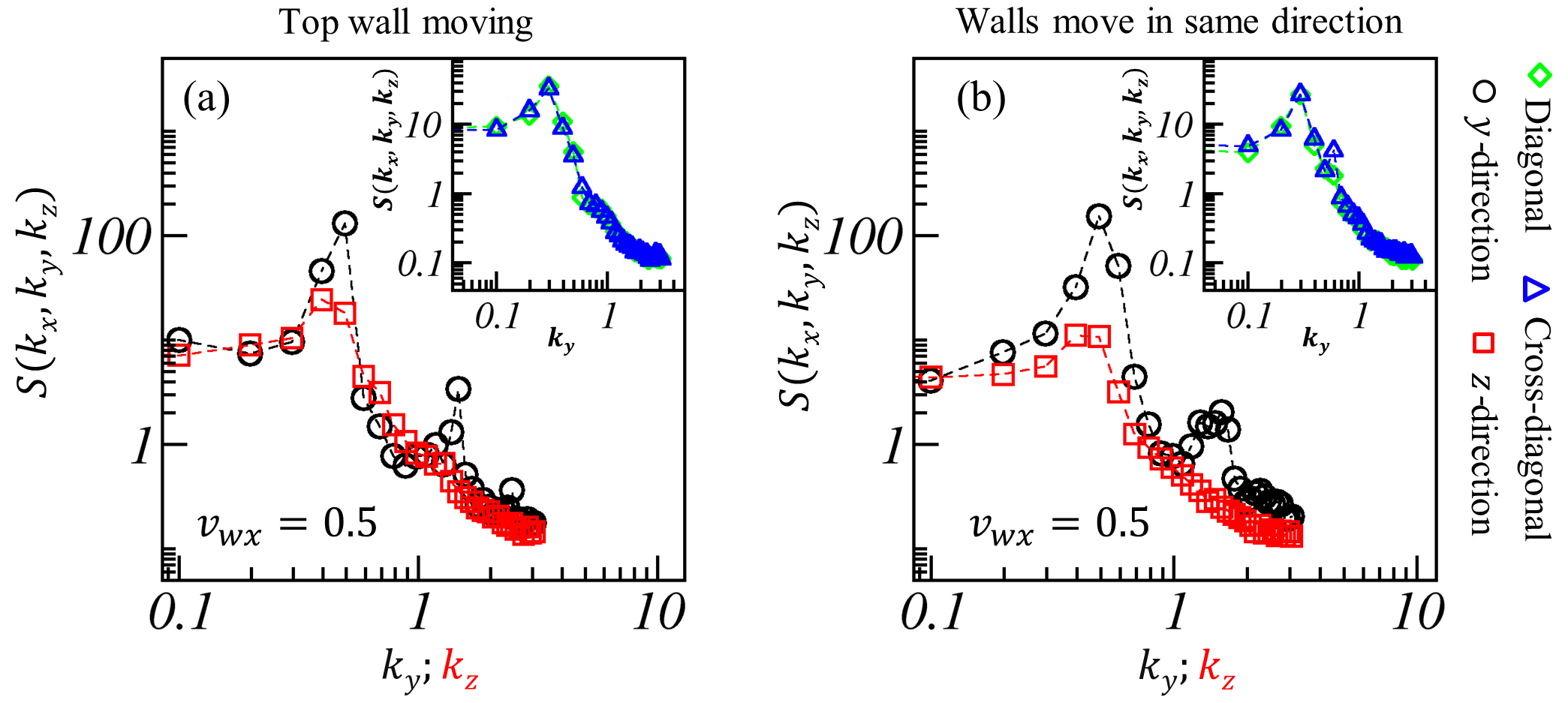}
	\caption{Plot of $S(k_x,k_y,k_z)$ versus $k_y$ (black curve) and $k_z$ (red curve) at $v_{wx}=0.5$ to illustrate the structural anisotropy (a) for the top wall motion along $x$-direction, and (b) when both walls move in same direction. The green and blue curves in the inset demonstrate $S(k_x,k_y,k_z)$ versus $k_y$ along diagonal and cross-diagonal directions respectively in $xz$-plane.}
	\label{fig7}
\end{figure}

Recall that $C(r,t)$ and $S(k,t)$ denote the spherically averaged correlation function and the structure factor. The critical BCP melts evolve to form lamellar morphology in the asymptotic limit. Thus, it is essential to manifest the anisotropy developed in the system at late times. In this regard, Figs. \ref{fig7}(a) and \ref{fig7}(b) display the structure factor, $S(k_x, k_y, k_z)$ against the wave vector components, $k_y$ (black curve) and $k_z$ (red curve) for case 2 and case 3 at $v_{wx}= 0.5$ for $t = 5400$. The black curve shows sharper and higher amplitude peaks in $S(k_x, k_y, k_z)$ along $k_y$ (normal to the lamellar). The red curve shows a single low amplitude peak along $k_z$ (parallel to the lamellar). Thus, both these curves justify the structural anisotropy in the system. The larger peak amplitude of the black curve for case 3 (Fig. \ref{fig7}(b)) explains a higher structure anisotropy than in case 2 (Fig. \ref{fig7}(a)). Notice the snapshots in Figs. \ref{SI2}-\ref{SI3}; they illustrate that lamellar slabs of each phase are coarsening nearly parallel to the $xz$-plane; thus, the variation of $S(k_x, k_y, k_z)$ along $k_x$ and $k_z$ is similar (comparison is not shown here due to brevity). However, excellent overlapping in green and blue curves in the insets of Fig. \ref{fig7} affirms the symmetric arrangement of lamellar slabs in the $yz$-plane. The $S(k_x,k_y,k_z)$ data is averaged over sixty-four ensembles. 

\subsection{Both walls move in opposite directions}
\label{bwm_od}
In case 4, wall motion generates equal and opposite shear in the $x$-direction. The evolution patterns in Fig. \ref{SI4} suggest that the system forms lamellar morphology much earlier than in cases 1-3. The reason could be the opposite shear directions facilitate a more rapid readjustment of evolving microdomains; it is evident that well-developed lamellar patterns are formed for $v_{wx}=0.5$ and $v_{wx}=1.0$. However, the system comprises a few shorter patches of $A$ and $B$ domains, elongated in the $x$-direction, at $v_{wx}=0.1$.

\begin{figure}[ht!]
	\centering
	\includegraphics[width=0.9\textwidth]{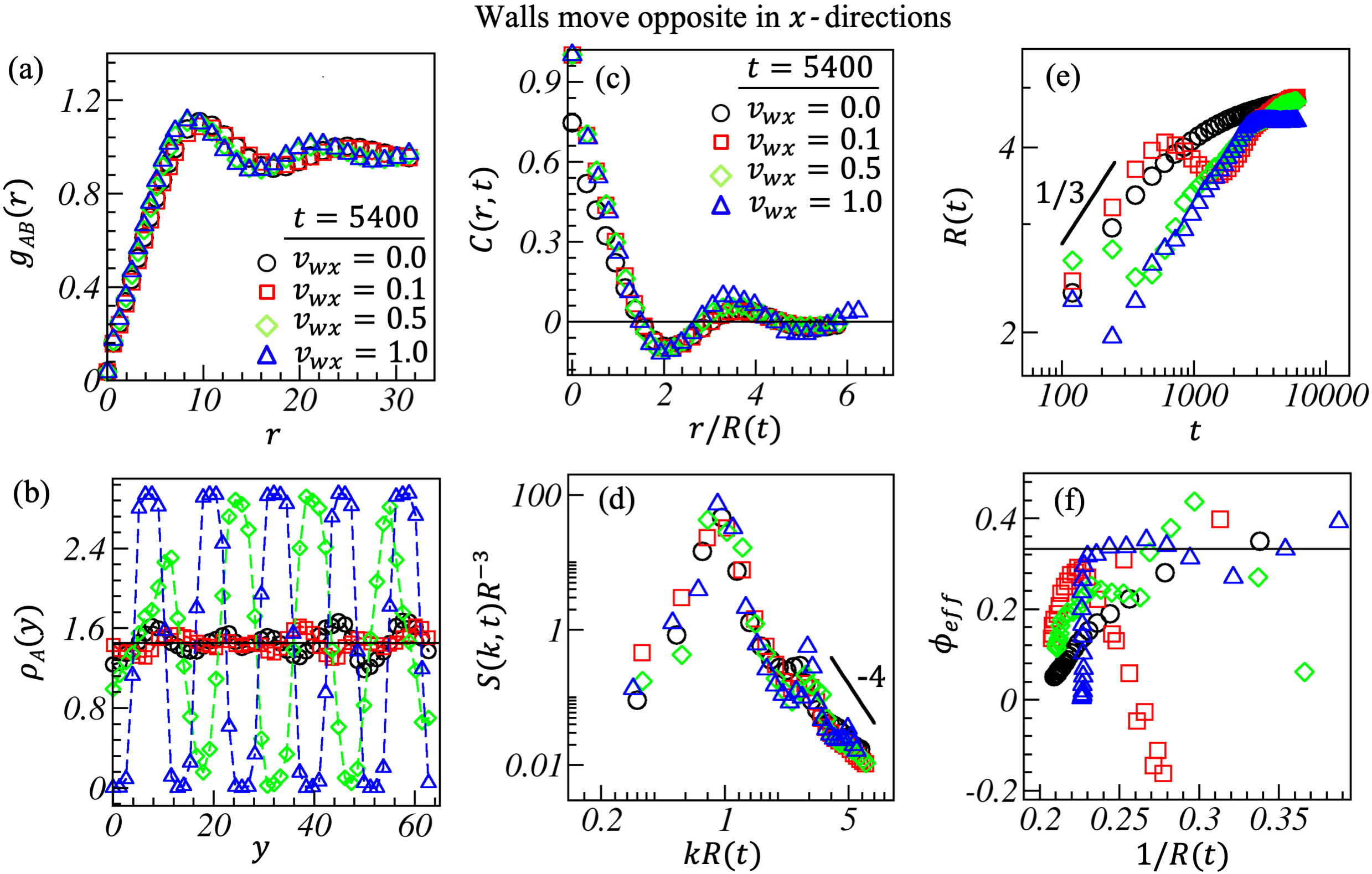}
	\caption{Case 4: both walls are moving opposite in $x$-direction with the same velocities. Comparison of (a) $g_{AB}(r)$ versus $r$, (b) $\rho(y)$ versus $y$ of $A$-type beads, (c) $C(r,t)$ versus $r/R(t)$, and (d) $S(k,t)$ versus $kR(t)$ at $t=5400$ for different wall velocities as shown with various open symbols. (e) The time dependence of average domain size $R(t)$ for the evolution is displayed in Fig. (\ref{SI4}). Corresponding effective growth exponents are plotted in (f).} 
	\label{fig8}
\end{figure}

As discussed earlier for other cases, $g_{AB}(r)$ vs. $r$ in Fig. \ref{fig8}(a) shows a similar late time behavior for all velocities. The peaks get more oscillatory and sharper with increasing shear rates due to the evolution of a more ordered and periodic lamellar morphology. A tiny shift in the first peak positions (at $r \simeq 9.5$, $9.0$, $8.32$) and a relatively larger change in secondary peak positions (at $r \simeq 24.96$, $22.40$, $21.76$) towards lower $r$ indicates the shear-thinning of morphologies with shear rate. Figure \ref{fig8}(b) illustrates that the amplitude of $\rho_A (y)$ vs. $y$ increases with the shear rate. A significant increase in the amplitude of $\rho (y)$ and much sharper domain interfaces signify a well-separated morphology at high shear strength ($v_{wx}= 1.0$), as displayed by the blue curve in Fig. \ref{fig8}(b). Nearly equal periodicity and amplitude of $\rho (y)$ profile along the $y$-direction manifests the formation of lamellar morphology normal to it at higher wall velocities; see the green and blue curves in Fig. \ref{fig8}(b) at $v_{wx}= 0.5$ and $1.0$.

The scaled $C(r, t)$ in Fig. \ref{fig8}(c) and the corresponding $S(k, t)$ in Fig. \ref{fig8}(d) demonstrate notable deviation from the dynamic scaling function in case 4. This is indifferent to case 2 and case 3, where we have observed the scaling in $C(r, t)$, at least up to the first zero crossing (at smaller $r$) and in $S(k, t)$ at larger $kR(t)$. Furthermore, $C(r, t)$ becomes more oscillatory at larger  $r/R(t)$ with the increase in $v_{wx}$. In $S(k, t)$, we observe prominent shoulders at larger $kR(t) \simeq 2.67$, $2.84$, $2.86$, and $2.93$ for $v_{wx}=0.0$, $0.1$, $0.5$, and $1.0$, respectively. In addition, however, we notice the development of another shoulder at $kR(t) \simeq 4.68$ and $4.95$ for $v_{wx} = 0.5$ and $1.0$. These results verify the formation of a more periodic lamellar morphology with an increasing shear rate than for other cases. Moreover, a fully evolved periodic lamellar morphology noted at $t=5400$ for $v_{wx}=1.0$.

The time evolution of characteristic length scale, $R(t)$ vs. $t$ in Fig. \ref{fig8}(e), and the corresponding growth exponent, $\phi_{eff}$ vs. $1/R(t)$ in Fig. \ref{fig8}(f) demonstrate the coarsening kinetics under the influence of shear for case 4. The black curve represents the reference case, $v_{wx}=0.0$. Similar to the earlier cases, for $v_{wx} = 0.1$, the length scale follows diffusive growth: $\phi \rightarrow 1/3$ up to $t_{sp} \simeq 600$ (red curve); the growth rate is higher within this period for $v_{wx} = 0.1$. Beyond this time, $R(t)$ deviates significantly. We find a gradual dip in the length scale up to $t\simeq 1400$ (see the red curve in Figs. \ref{fig8}(e-f); the reason could be the domination of shear rate over the diffusive growth, which is already slowing down due to the topological constraint of BCP chains. Therefore, as noted earlier, BCP chains in microdomains are possibly rearranging perpendicular to the shear direction to reduce its effect may cause the dip during this period. The green and blue curves at $v_{wx} = 0.5$ and $1.0$ also illustrate the similar dip in growth at much earlier times. After the dip in growth, $R(t)$ begins to grow further and in contrast to cases 1-3, it follows diffusive growth ($\phi \sim 1/3$) for all $v_{wx}\neq 0.0$ for a longer period before saturating. The length scale fully saturates at a lower $R(t)$ for $v_{wx} = 1.0$ (see the blue curves in Figs. \ref{fig8}(e-f)), which is evident from Fig. \ref{SI4}(c) exhibiting a well-formed lamellar morphology. However, the length scales for $v_{wx} = 0.1$ and $0.5$ seem to saturate at the black curve (reference case) at late times. The effective growth exponent in  Fig. \ref{fig8}(f) demonstrates the saturation in length scale in the asymptotic limit. Overall, the effect of shear, rendered by the horizontal parallel walls moving in opposite directions, is significant in obtaining the typical lamellar morphology much earlier in the microphase separating critical BCP melt.

\begin{figure}[ht!]
	\centering
	\includegraphics[width=0.7\textwidth]{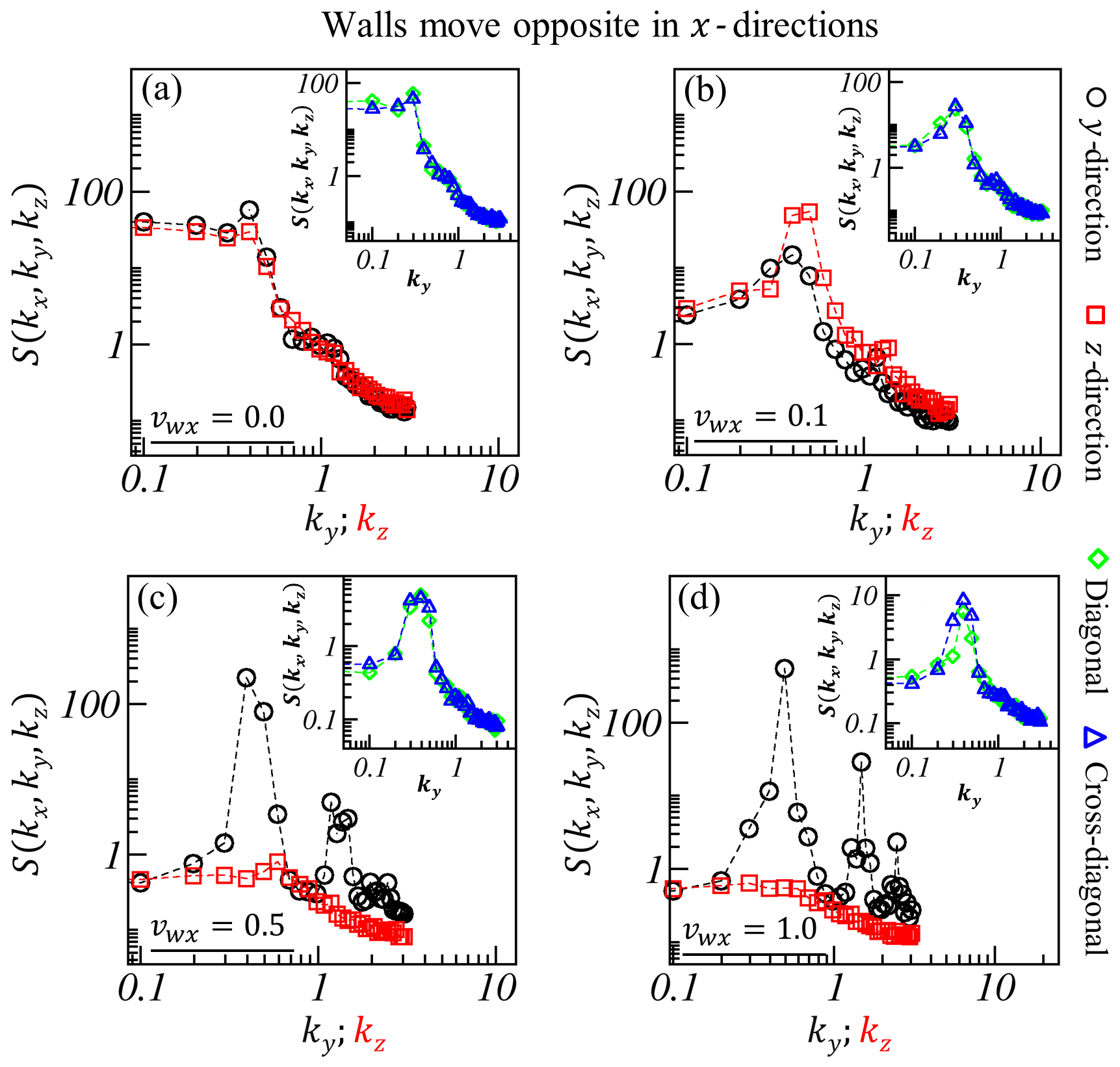}
	\caption{The structure factor, $S(k_x,k_y,k_z)$ versus $k_y$ (black curve) and $k_z$ (red curve) for (a) $v_{wx}=0.0$, (b) $v_{wx}=0.1$, (c) $v_{wx}=0.5$, and (d) $v_{wx}=1.0$ to display the structural anisotropy in the system. The green and blue curves in the inset demonstrate $S(k_x,k_y,k_z)$ versus $k_y$ along diagonal and cross-diagonal directions respectively.}
	\label{fig9}
\end{figure}

Recall that non-overlapping of the structure factor, $S(k_x,k_y,k_z)$  along different directions, $k_x$, $k_y$, and $k_z$ represent the structural anisotropy in the system. Here, we discuss the anisotropy in the evolution morphology at a late time $t = 5400$ for $v_{wx} = 0.0$, $0.1$, $0.5$, $1.0$, as displayed in Figs. \ref{fig9}(a-d). The black and red curves represent the variation of $S(k_x,k_y,k_z)$ vs. $k_y$ (normal to the stripes) and $k_z$ (along with the stripes). For $v_{wx}=0.0$, lamellar patterns are not fully developed in any particular direction. However, $S(k_x,k_y,k_z)$ data along $k_y$ (black curve) shows a slightly larger peak and also a developing shoulder at larger $k_y$ than in the red curve; this indicates that lamellar patterns may evolve normal to the $y$-direction at much longer times. Further, notice the stripes for $v_{wx}=0.1$ in Figs. \ref{SI4}(a), which are nearly parallel to the $xy$-plane (i.e., normal to the $z$-axis). Thus, we notice a moderately higher amplitude of structure factor peak at a smaller $k_z$ (red curve) than in the black curve against $k_y$. However, an emerging shoulder is apparent in both curves for the large $k_y$ and $k_z$.

As noted earlier from Figs. \ref{SI4}(b-c), more distinct lamellar stripes are formed at higher shear rates, i.e., at  $v_{wx} = 0.5$ and $1.0$, normal to the $yz$-plane along the $x$-direction. Therefore, we notice multiple sharp, high amplitude peaks in the structure factor along the $y$-direction (see the black curves in Figs.\ref{fig9}(c-d); these peaks depict the periodic $A$ and $B$-type stripes. In contrast, the structure factor peak seen in the red curve at lower shear rates diminishes at $v_{wx} = 0.5$ and $1.0$. The reason is that the stripes get more and more parallel to the $xz$-plane with the increase in shear rate along the $x$-axis. Hence, these results verify the presence of anisotropy in the $y$-direction. Furthermore, the green and blue curves in the inset of Figs.\ref{fig9}(a-d) exhibit the variation in $S(k_x,k_y,k_z)$ along the diagonals in the $yz$-plane versus $k_y$. The excellent data overlap for all shear rates justifies the structural symmetry in the $yz$-plane along the diagonals. Also, notice the structure factor peaks shown by the green and blue curves; the main peaks get sharper and higher in amplitude with enhanced shear, indicating more ordered lamellar morphology. We average the $S(k_x,k_y,k_z)$ data over sixty-four ensembles. 

\begin{figure}[ht!]
	\centering
	\includegraphics[width=0.9\textwidth]{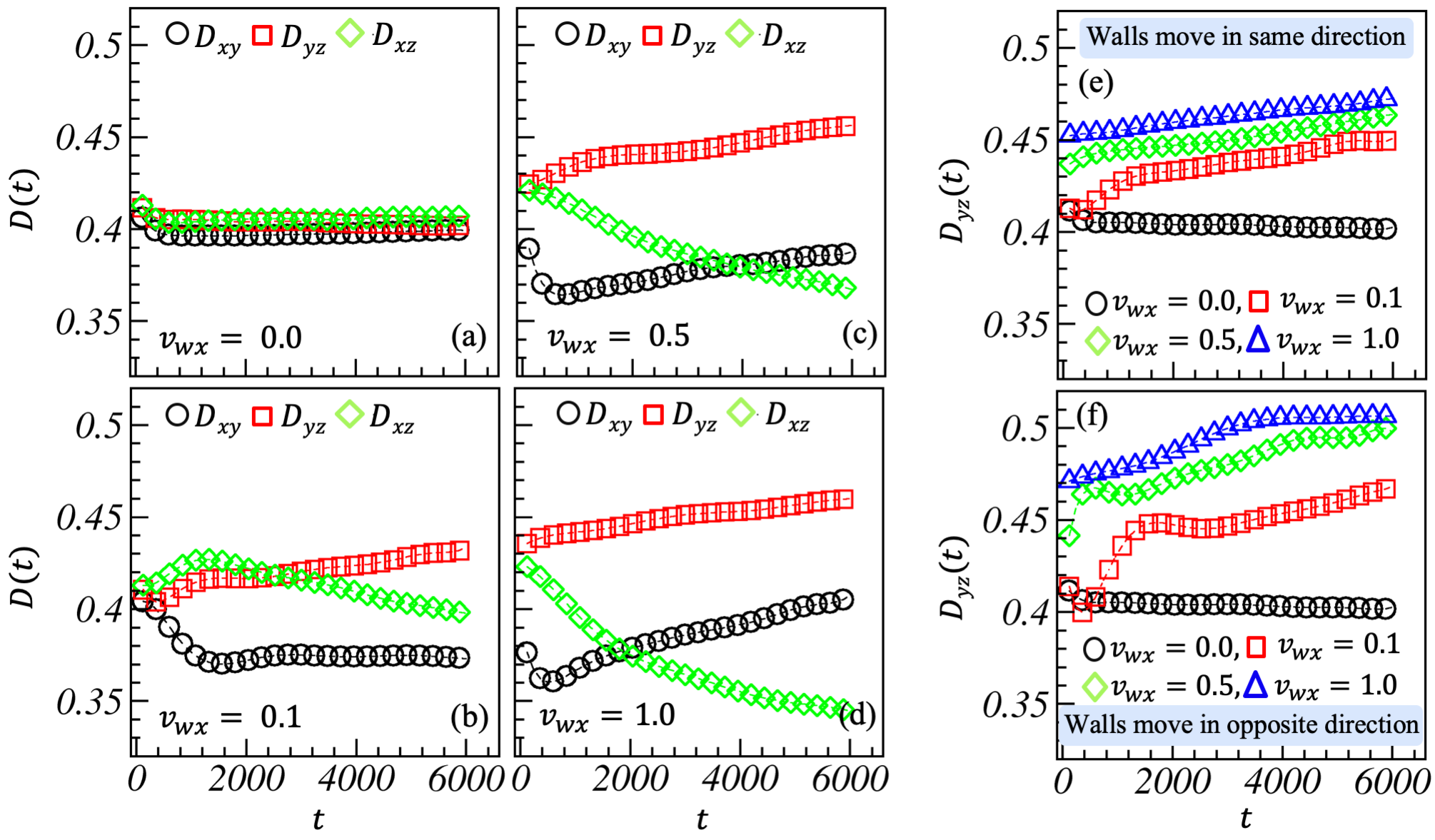}
	\caption{Anisotropy parameters, $D_{xy}, D_{yz}$, and $D_{xz}$ are plotted as a function of time for case 1 ($v_{wx} = 0.0$) in (a) and for case 2 in (b-d) for different wall velocities, $v_{wx} = 0.1$, $0.5$, and $1.0$. Time variations of $D_{yz}$ are compared in (e) and (f) for case 3 and case 4 at different wall velocities depicted by various symbol types.}
	\label{fig10}
\end{figure}

The time variation of the anisotropy parameter, $D(t)$, is plotted in Fig. \ref{fig10} as denoted by various symbol types. Usually, due to structural constraints, phase separating BCP melts have inherent anisotropy. Therefore, for case 1, we have $D(t)\sim 0.4$ (see Fig. \ref{fig10}(a)), i.e., $D_{xy}$, $D_{yz}$, and $D_{xz}$ values are approximately same in this case. Thus, at our simulation's time and length scale, the typical lamellar morphology for a symmetric BCP melt is not fully evolved yet for case 1 (reference case, see Figs. \ref{fig1}(a) and \ref{SI1}). For case 2, due to top wall shear in the $x$-direction, microdomains begin to reorient themselves in forming lamellar slabs parallel to the $xz$-plane (as in Fig. \ref{SI2}). As a result, $D(t)$ decreases in the $xz$-plane (green curve) and grows $yz$-planes (red curve) with time. Furthermore, we notice that the differences in $D(t)$ in different planes become more prominent at late times, particularly for higher wall velocities. Thus, with the increase in wall velocity, anisotropy in the $x$-direction (i.e., in the $yz$-plane) increases more due to more ordered evolved morphology and decreases in the $xz$-plane. At a low shear rate, $v_{wx} = 0.1$, a decrease in $D_{xy}$ (black curve in Fig. \ref{fig10}(b)) suggests that only small and local slabs of both the phases are developed in the $xy$-plane without any specific orientation. However, $D_{xy}$ increases with higher shear rates. In Fig. \ref{fig10}(e) and (f), we compare the time variation of $D_{yz}$ for case 3 and case 4 at different shear rates. In both cases, we observe that evolved morphologies are more anisotropic compared to case 2, as depicted in Fig. \ref{fig10}(b)-(d). Nevertheless, as discussed earlier, we find a clearer lamellar morphology for case 4; thus, the corresponding anisotropy parameter, $D_{yz}$, is also relatively higher than in cases 2 and 3. The corresponding evolution snapshots are presented in Figs. \ref{SI2}, \ref{SI3} and \ref{SI4}, respectively.

\begin{figure}[ht!]
	\centering
	\includegraphics[width=0.8\textwidth]{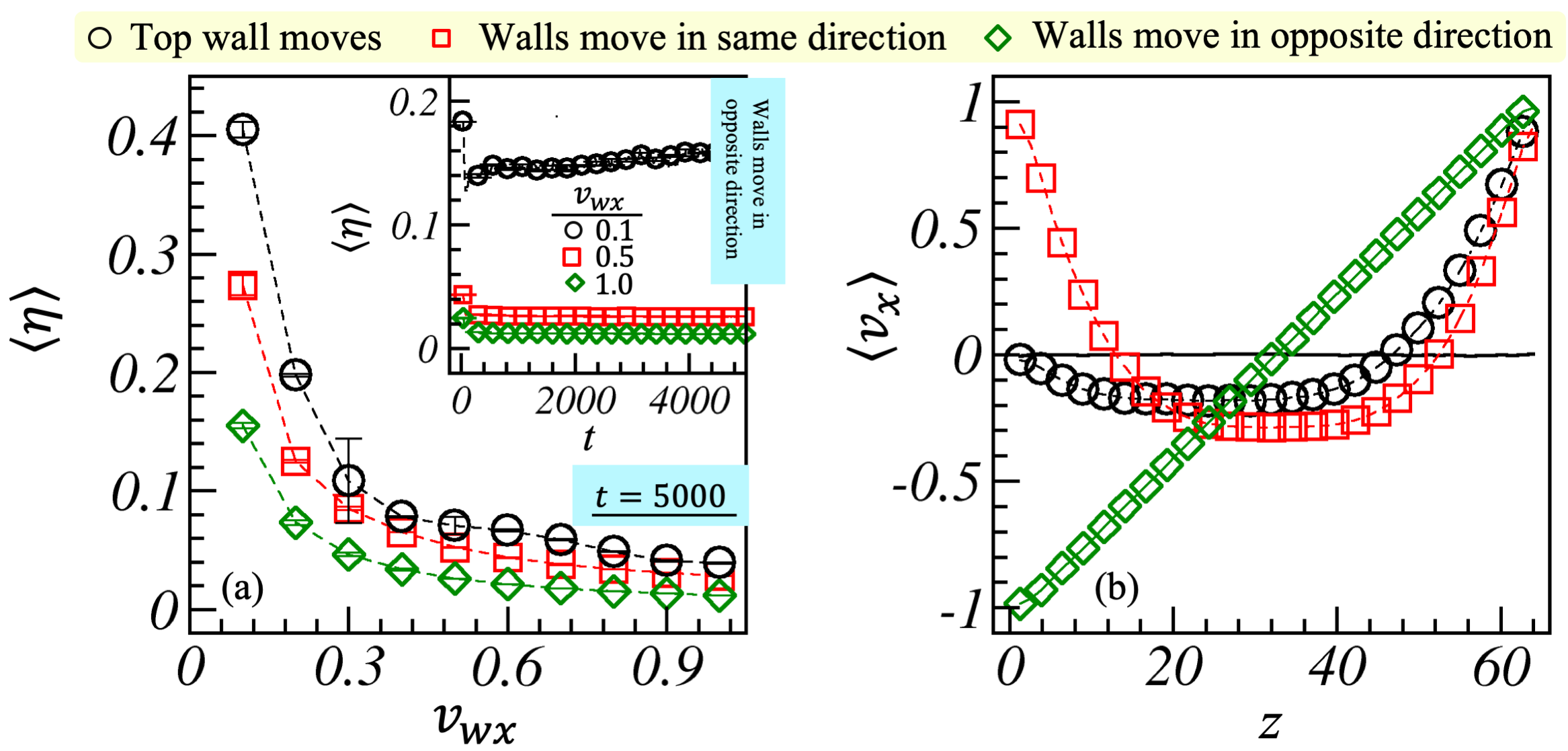}
	\caption{(a) The variation of shear viscosity, $\langle \eta\rangle$ against $v_{wx}$ at $t=5000$. The different symbol types at the top represent different cases studied here. The inset depicts the time variation of $\langle \eta\rangle$ for case 4 at different wall velocities indicated by various symbols in the legend. (b) Average velocity profile, $\langle v_x\rangle$ is plotted along the $z$-direction at $v_{wx}=1.0$ for all the cases.}
	\label{fig11}
\end{figure}

Finally, to understand the effect of shear on the phase separating BCP melt in bulk, we plot the variation of shear viscosity $\langle\eta\rangle$ with $v_{wx}$ in Fig. \ref{fig11}(a) for cases $2$-$4$ as denoted by the black, red, and green symbols at a late time of evolution, $t = 5000$. The symbol $\langle \cdots \rangle$ denotes the average over ten ensembles. For each case, the variation of $\langle\eta\rangle$ follows a similar pattern. i.e., with the increase of shear rate, a gradual decrease in viscosity is noted, illustrating the shear-thinning, which further saturates at higher shear rates ($v_{wx}$). Thus confirming the smaller average domain size, $R(t)$ at higher $v_{wx}$ for cases 2-4 discussed above. The shear viscosity for case 4 (green curve) is the lowest, indicating that the effect of shear is more prominent in this case than the cases 2 and 3. In the inset of Fig. \ref{fig11}(a), we illustrate the variation of $\langle\eta\rangle$ as a function of time at $v_{wx}\neq 0$ depicted by various symbol types for case 4. After an early decrease in shear viscosity, it saturates quickly to an average value with tiny statistical fluctuations. However, for a higher shear rate, we observed lower shear viscosity.

Since the walls move along the $x$-direction, $v_x$ plays a dominant role in the velocity profile. Therefore, we generate the velocity profiles, $\langle v_x \rangle$, along the $z$-direction for all the cases at $v_{wx}= 1.0$, as displayed in Fig. \ref{fig11}(b). For this, we divide the system into thin slabs of unit thickness along the $z$-direction and calculate the average velocity components $\langle v_x \rangle$, $\langle v_y \rangle$, and $\langle v_z \rangle$ over all the beads in a particular slab.\cite{muller1999reversing,kelkar2007prediction} The $y$ and $z$ components of the average velocity profile remain unaffected by the shear; thus, we get $\langle v_y \rangle = \langle v_z \rangle = 0$ for all the cases (results are not shown here). The solid black line demonstrates the velocity profile for case 1, where both walls are fixed, and hence, $\langle v_x \rangle = 0$. This is plotted as a reference curve for the other cases studied in Fig. \ref{fig11}(b) at $v_{wx}=1.0$. In case $2$ (black symbol), as the distance from the top wall increases, $\langle v_x \rangle$ gradually reduces to zero. It reduces further and attains negative values with distance, which gradually approaches zero close to the bottom wall. The velocity profile for case $3$ (red curve) decreases smoothly with increasing distance from both the walls and reaches a negative minimum value at $z\simeq L/2$. Finally, in case $4$ (green curve), since both walls are moving in opposite directions, $\langle v_x \rangle$ varies linearly with distance along the $z$-axis by crossing zero near $z \simeq L/2$. 

Overall, the symmetric variation of the velocity profile within positive and negative values (i.e., the momentum flow in opposite directions) ensures momentum conservation in the system. The momentum exchange further ensures the drainage of excess heat generated by friction.\cite{muller1999reversing} Thus, we observe only tiny enhancements in the instantaneous temperature, $T_i \in (0.98 - 1.095)$ around the quench temperature, $T = 1.0$, but within the statistical error at various shear rates in different cases, as displayed in Fig. \ref{SI5}.

\section{Summary and conclusions}
\label{sumcon}
We utilize the DPD simulation technique to simulate the phase separation kinetics of the critical BCP bulk melts under three unidirectional shear conditions. The shear is applied by moving the parallel solid walls with a constant velocity in a particular direction only. The walls confine the cubic simulation box at the top and the bottom. The shear is allowed to affect the system in three ways: (i) only the top wall moves (case 2). (ii) Both walls move in the same direction (case 3). (iii) In the last case, both walls move in opposite directions (case 4). We discuss the effect of shear rendered in the system on evolution morphology, scaling functions, length scale, and growth law for different cases. Further, we compare the anisotropy induced in the system from the structure factor variation in different directions and justify it further by computing the anisotropy parameter. All the shear-influenced results are compared with no shear case, \textit{i.e.}, when both the walls are fixed (case 1). Finally, we present the variation in shear viscosity and velocity profiles for all the cases with wall velocity, which is proportional to the shear rate.

On the application of shear, microdomains start flowing along the shear direction. We notice that the lamellar formation always begins in the region close to the moving wall. One of our significant observations is that segregating BCP chains, regardless of chain size, begin to adjust perpendicular to the shear direction and along the wall plane to overcome its effect at moderate to high shear rates. This eventually leads to a periodic lamellar pattern much earlier than case 1 (without shear). However, a fully grown lamellar morphology is still not observed at the lower shear rate up to the simulation period used in this work; nonetheless, our results suggest they are more periodic than in case 1. The corresponding radial distribution function, the local density profile variation, and the comparison of structure factors in different directions perfectly characterized the morphology under various shear conditions. The late time correlation function and the structure factor offer a reasonably good scaling for all the shear rates in case 2 and case 3. However, in case 4, where the shear effect is more pronounced on the evolution morphology, the scaled correlation function and the structure factor deviated from the scaling for all the velocities. The oscillatory nature of the correlation function and the emerging shoulder at larger $kR(t)$ values in the structure factor further justify the segregating periodic morphology. Moreover, broadening of the structure factor main peak at various shear rates signifies the shear-thinning of domain size in the asymptotic limit. 

The competition between the microphase separation kinetics and the shearing through wall motion leads to some fascinating behavior of characteristic length scale. In particular, at lower shear rates, the flow of early small domains and individual BCP chains in the shear direction seems to complement the domain segregation. As a result, the average domain size becomes more prominent for all the cases at early times. The variation of average shear viscosity with shear rates (wall velocities) justifies that at lower shear rates, shear viscosity is higher (i.e., low shear-thinning) than at moderate and high shear rates. Interestingly, near the length scale crossover to saturation for case 1, the length scale at the lower shear rates also slows significantly (the length scale became almost flat) for case 2 and case 3 within a specific time interval.

On the other hand, we observe negative growth for case 4 for a slightly larger period due to the shear effect dominating the diffusive growth during this time interval. As a result, the BCP chains are merely adjusting themselves to minimize the shear effect. Again, the length scale grows with time following the diffusive growth and then begins to saturate at a length scale reasonably lower or nearly equal to case 1 at late times for different cases. Furthermore, moderate and large shear rates dominate the diffusive growth from the beginning for almost all cases. Thus, the length scale growth rate, lower from early times, saturates to a lower length scale at late times, indicating the shear-thinning. Again, the variation of average shear viscosity with shear rates justifies the same.

Overall, this work illustrates the influence of unidirectional shear induced by moving solid walls on phase separation kinetics of the critical BCP bulk melts. Besides, this work can guide quickly producing bulk composite materials with diverse anisotropic morphology and various physical properties. Given the different scientific and technological importance of morphologies obtained from segregating BCP melts, these results will enable additional experimental and theoretical interest in this problem. Further, we hope this work will offer a broad framework to simulate the morphology growth in BCP melts and polymer blends under such an external response.
 
\section*{Conflicts of interest}
There are no conflicts of interest to declare.
\section*{Author Contributions}
Conceptualization, AS, and AKS.; Methodology, AKS; Validation, AS; Formal analysis and Investigation, AKS, and AS; Resources and Data curation, AKS; Writing-original draft preparation, AKS; Writing-review and editing, AKS; Visualization, AKS; Supervision, AS; Funding acquisition, AS and AKS. All authors have read and agreed to the published version of the manuscript.
\section*{Acknowledgement}
A.K.S. is grateful to the CSIR fellowship (File No: 09/1217(0074)/2019-EMR-I), New Delhi, for the financial support. A.S. acknowledges the financial support from Grant No. ECR/2017/002529 by the Science and Engineering Research Board, Department of Science and Technology, New Delhi, Government of India.
\section*{Data Availability}
Data will be made available on request.

\bibliographystyle{unsrt}
\bibliography{mybib}

\newpage
\section*{Supplementary Information}
\setcounter{figure}{0}
\renewcommand{\thefigure}{S\arabic{figure}}
\renewcommand{\figurename}{Figure}
\begin{figure}[ht!]
\centering
\includegraphics[width=0.75\textwidth]{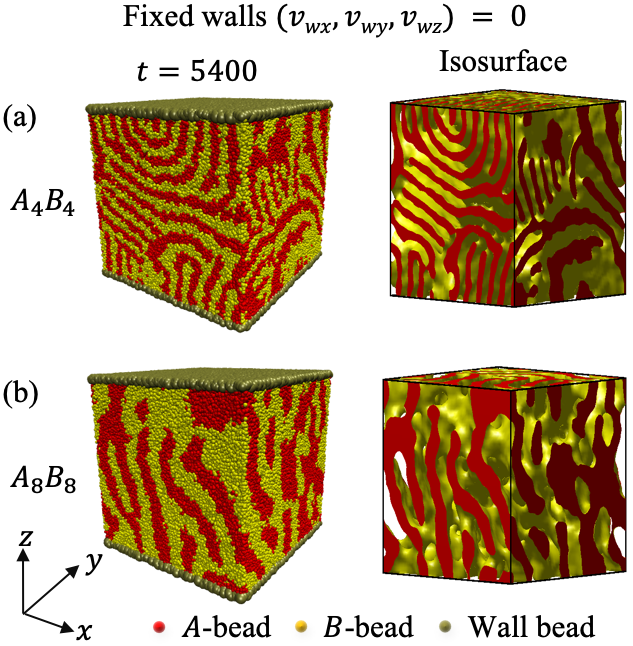}
\caption{Comparison of late stage morphology of phase separating critical BCP melt for (a) $N_p=8$ and (b) $N_p=16$. The second right frame shows the corresponding isosurfaces.}
\label{SI1}
\end{figure}
\clearpage

\begin{figure}[ht!]
\centering
\includegraphics[width=0.9\textwidth]{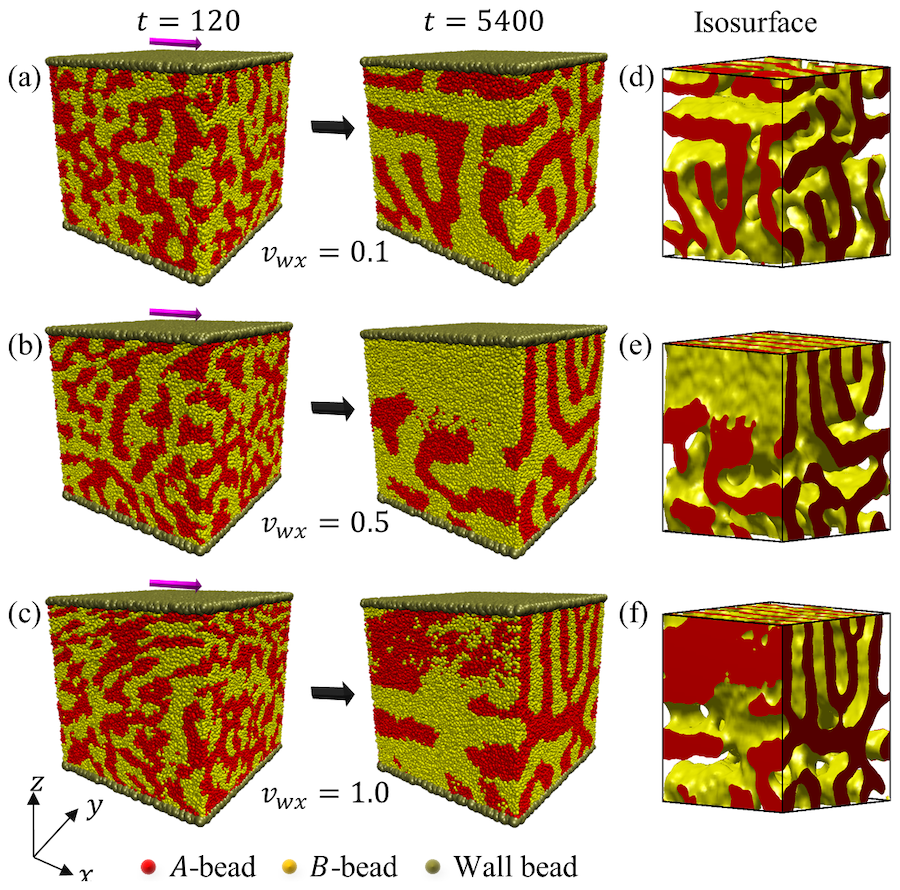}
\caption{Morphology evolution of BCP melt when the top wall is moving in positive $x$-direction with different velocities (a) $v_{wx} = 0.1$, (b) $v_{wx} = 0.5$, and (c) $v_{wx} = 1.0$ . The first column snapshots are at an early time ($t=120$), and the second column images are at a late time step ($t=5400$). The magenta arrows show the shear direction. The third column displays the isosurfaces for the snapshots in the second column.}
\label{SI2}
\end{figure}
\clearpage

\begin{figure}[ht!]
\centering 
\includegraphics[width=0.9\textwidth]{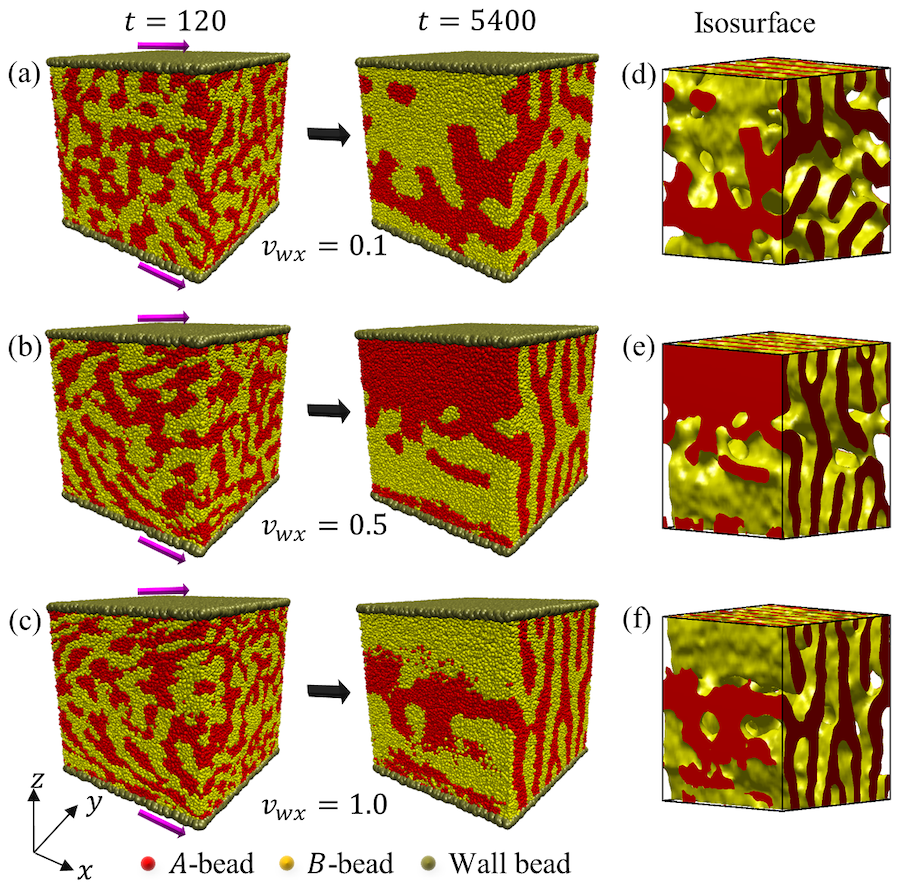}
\caption{The evolution snapshots at two time steps $t=120$ and $t=5400$ when top and bottom walls are moving in the same direction with the same velocities (a) $v_{wx}=0.1$, (b) $v_{wx}=0.5$, and (c) $v_{wx}=1.0$, respectively. The third column displays the isosurfaces for the snapshots in the second column. The magenta arrows show the shear direction.}
\label{SI3}
\end{figure}
\clearpage

\begin{figure}[ht!]
\centering
\includegraphics[width=0.9\textwidth]{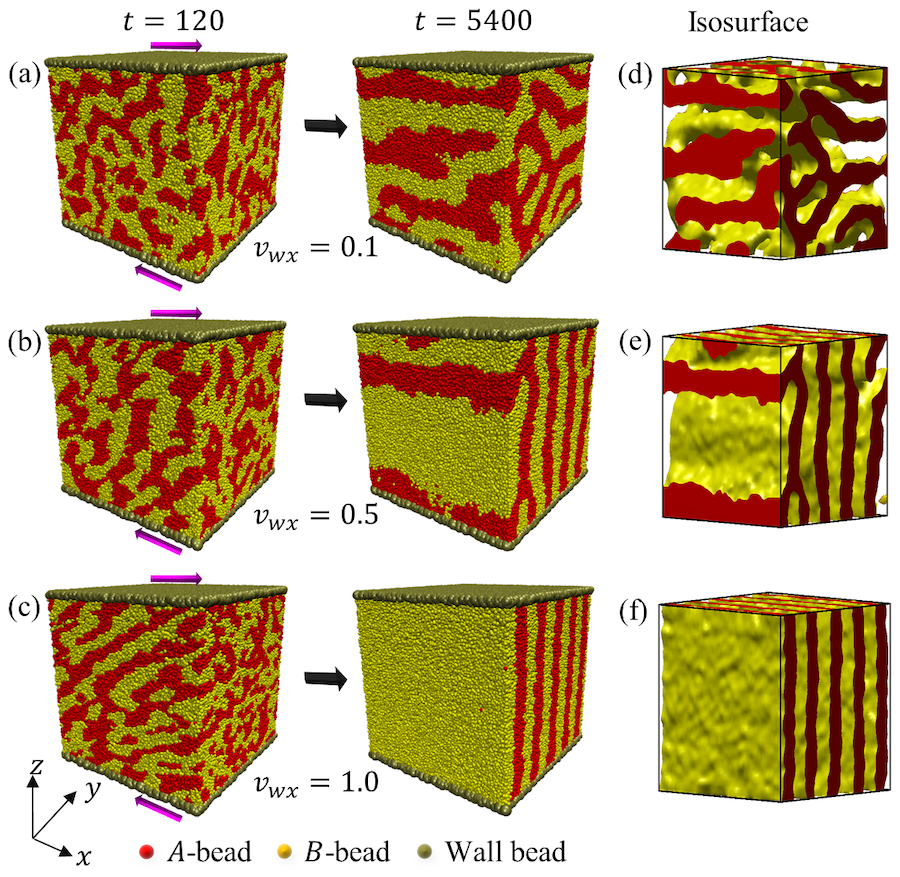}
\caption{Evolution morphologies at two-time steps for wall velocities (a) $v_{wx}=0.1$, (b) $v_{wx}=0.5$, and (c) $v_{wx}=1.0$ when both walls are moving with the same velocity but in the opposite direction; the top wall is moving in positive $x$-direction and the bottom wall is moving in negative $x$-direction. The third column displays the isosurfaces for the snapshots in the second column. The magenta arrows show the shear direction.}
\label{SI4}
\end{figure}
\clearpage

\begin{figure}[ht!]
\centering
\includegraphics[width=0.8\textwidth]{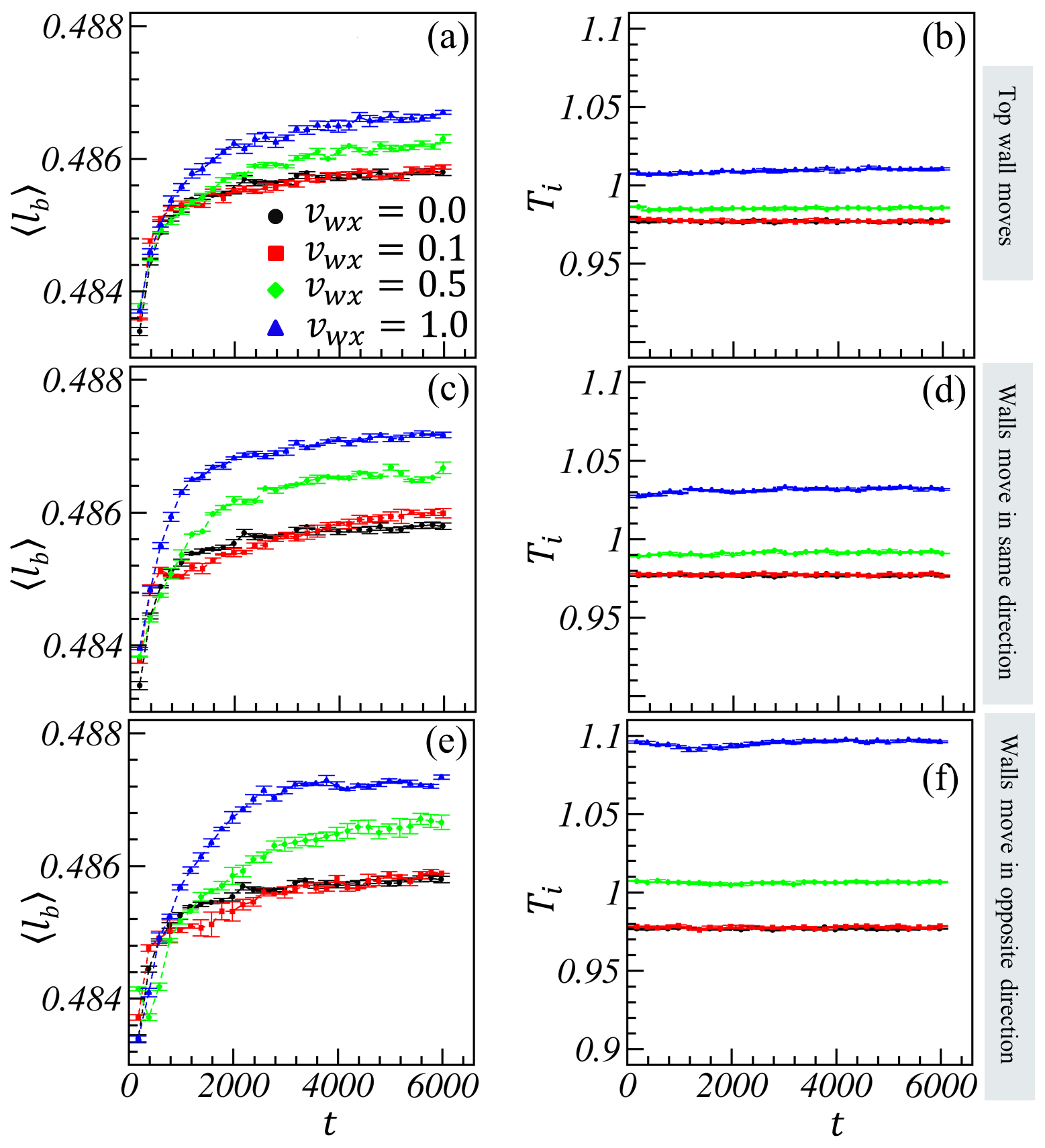}
\caption{The first column ((a), (c), and (e)) compares the time variation of average bond length, $\langle l_b\rangle$ at different shear rates, $\dot{\gamma} \propto v_{wx}$, as denoted by various symbols in the legend for three cases displayed at the extreme right. These curves illustrate the negligible bond stretching with increased shear rates caused by wall velocities. The second column ((b), (d), and (f)) compares the instantaneous temperature variation with time for the same cases as in the first column. Again, we observe tiny enhancements in $T_i \in (0.98 - 1.095)$ around the quench temperature, $T = 1.0$, but within the statistical error at various shear rates in different cases. In short, these results certainly disapprove any ambiguity that may cause due to the high shear rates, mainly due to excess bond stretching or temperature enhancement.}
\label{SI5}
\end{figure}
\clearpage

\end{document}